\documentclass[journal]{IEEEtran}
\ifCLASSINFOpdf
\else
\fi
\usepackage{mathtools}
\usepackage{amsmath}
\usepackage{graphicx}
\usepackage{amsmath}
\usepackage{amssymb}\usepackage[noend]{algpseudocode}
\usepackage{algorithmicx}
\usepackage{verbatim}
\usepackage{subcaption}
\usepackage{caption}
\usepackage{hyperref}
\usepackage{lipsum,amsmath,multicol}
\usepackage{color,soul}
\usepackage{qtree}
\usepackage{xcolor}
\usepackage{amsthm}
\usepackage{algorithm}
\usepackage{algpseudocode}
\usepackage{pifont}
\usepackage{url}

\begin{document}
%

\title{Cache-Aware QoE-Traffic Optimization in Mobile Edge Assisted Adaptive Video Streaming}

%
%
%

\author{Abbas Mehrabi,~\IEEEmembership{Member,~IEEE,}
        Matti Siekkinen,~\IEEEmembership{Member,~IEEE,}
        Antti Yl{\"a}-J{\"a}{\"a}ski,~\IEEEmembership{Member,~IEEE}
\thanks{Abbas Mehrabi, Matti Siekkinen and Antti Yl{\"a}-J{\"a}{\"a}ski are with the Department of Computer Science, Aalto University, Espoo, Finland. P.O.Box 15400 FI-00076.  \protect\\ 
E-mail: \{abbas.mehrabidavoodabadi, matti.siekkinen, antti.yla-jaaski\}@aalto.fi}}

%
%

\markboth{IEEE/ACM Transactions on Networking}  
{Shell \MakeLowercase{\textit{et al.}}: Bare Demo of IEEEtran.cls for IEEE Journals}
%



\maketitle

\begin{abstract}
Multi-access edge computing (MEC) enables placing video content at the edge of the network  
aiming to improve the quality of experience (QoE) of the mobile clients. 
Video content caching at edge servers also reduces traffic in the backhaul of the mobile network, hence reducing operational costs for mobile network operators (MNOs). 
However, minimizing the rate of cache misses and maximizing the average video quality may sometimes be at odds with each other, particularly when the cache size is constrained. Our objective in this article is two fold: First, we explore the impact of fixed video content caching on the optimal QoE of mobile clients in a setup where servers at mobile network edge handle bitrate selection. Second, we want to investigate the effect of cache replacement on QoE-traffic trade-off. An integer nonlinear programming (INLP) optimization model is formulated for the problem of jointly maximizing the QoE, the fairness as well as minimizing overall data traffic on the origin video server. Due to its NP-Hardness, we then present a low complexity greedy-based algorithm with minimum need for parameter tuning which can be easily deployed.    
We show through simulations that the joint optimization indeed enables striking a desired trade-off between traffic reduction and QoE. The results also reveal that with fixed cached contents, the impact of caching on the QoE is proportional to the desired operational point of MNO. Furthermore, the effect of cache replacement on QoE is less noticeable compared to its effect on backhaul traffic when cache size is constrained.  
\end{abstract}

\begin{IEEEkeywords}
Multi-access edge computing (MEC), Video caching, Adaptive video streaming, QoE, Fairness, Integer nonlinear programming (INLP), Greedy-based algorithm.   
\end{IEEEkeywords}

%
\IEEEpeerreviewmaketitle

\section{Introduction}
\label{sec:introduction}

\IEEEPARstart{A}{ccording} to statistics, the majority of Internet traffic is generated by video streaming applications, such as Netflix, YouTube etc.~\cite{Sandvine}. It is expected that over 70\% of the Internet traffic will be generated by video streaming applications by 2019 \cite{Tran2017}. The low end-to-end communication latency and high available bandwidth of future 5G mobile networks will enable mobile clients to view much higher quality video content than today~\cite{Tran2017}. In 5G networks, Multi-access Edge Computing (MEC), where servers are deployed within the radio access network (RAN)~\cite{Taleb2017}, and Network Function Virtualization (NFV) provide the platform to serve content from the edge of the network, hence reducing contention on the backhaul network~\cite{Liang2015,Tran2017}.

On the other side, the dynamic network conditions, namely fluctuation of the available bandwidth when multiple clients simultaneously compete for it and dynamic radio link conditions due to user mobility, can significantly affect the quality of experience (QoE) in mobile video streaming~\cite{Yao2011,Riiser2012}. Adaptive streaming protocols such as the non-standard HTTP Live Streaming (HLS) or protocols based on the dynamic adaptive video streaming (DASH) standard~\cite{DASH} can adapt to dynamic network conditions. In adaptive streaming, the whole video is divided into chunks and encoded into multiple qualities on the server~\cite{Seufert2015}. The client adapts dynamically to bandwidth fluctuations by downloading the most sustainable bitrate for each chunk of video and thereby striving to maximize its overall QoE. 

Content caching and delivery at the mobile edge servers provides means for 
Internet service providers (ISPs) to substantially cut their costs by reducing backhaul network and inter-ISP traffic~\cite{Wang2014}. In addition to reducing traffic on the backhaul network, edge caching of video content can also improve client perceived QoE through reduced latency and possibility of network congestion. Although caching has been thoroughly studied in different contexts, we believe that the interplay of adaptive video streaming and edge caching has not received yet sufficient attention. 

The problem we tackle stems from edge caching combined with regular adaptive video streaming client, using HLS or DASH, which only try to maximize their own video bitrate given bandwidth constraints. This combination causes cache misses when individual clients choose different quality chunks even if they are streaming the same video content, in particular when the cache size is relatively small. The goal of our work is to reduce the rate of cache misses by designing a coordinated bitrate selection strategy that is aware of the cache, hence making clients prefer to choose chunks that are already cached even if it leads to slightly lower average bitrate. In this way, our solution enables striking a desired trade-off between traffic reduction due to increased cache hit rate and QoE. We design this trade-off to be parameter controlled so that ISPs can tune it to meet their needs. What makes the solution attractive is that in some situations, small sacrifice in video quality leads to significant reduction in backhaul and/or inter-ISP traffic.



Our main contributions in this work are summarized as follows: 
\begin{itemize}
\item We examine joint optimization of QoE and data traffic in mobile edge assisted adaptive video streaming where the video contents can be cached at the edges of the network. 
\item Aiming to quantify the impact of content caching on the optimal QoE of the mobile clients, we propose an INLP optimization problem for jointly maximizing the QoE and fairness as well as minimizing the data traffic on the origin video server. 
\item Due to NP-hardness of the problem formulation, a greedy-based scheduling algorithm with low complexity is then designed. It has the minimum need for parameter tuning which can be easily deployed by mobile network operators (MNOs). 
\item We discover through simulation that with fixed contents cached at the edges, the caching significantly impacts the QoE of the clients but it can be parameter controlled with the joint optimization. 
Furthermore, the impact of cache updating heuristic on QoE is less noticeable compared to its impact on data traffic when the cache size is constrained.   
\end{itemize}

The rest of the paper is organized as follows: We discuss related work in Section \ref{sec:relatedwork} and describe the mobile edge caching adaptive video streaming design together with its components in Section \ref{sec:framework}. The joint QoE-traffic optimization problem is laid out in Section \ref{sec:optimizationproblem} and the proposed centralized scheduling algorithm, the cache updating heuristic and the complexity analysis are detailed in Section \ref{sec:algorithm}. We present simulation-based evaluation in Section \ref{sec:results} and finally conclude the paper in Section \ref{sec:conclusion}.

\section{Related Work}
\label{sec:relatedwork}

Multi-access edge computing (MEC) concept proposed by European Telecommunication Standard Institute (ETSI) is a key solution toward reducing the contention in the backhaul network~\cite{Tran2017,Taleb2017}. \emph{Wang et. al} \cite{Wang2017} provide a comprehensive survey on the advantages of MEC architecture compared to centralized cloud-based framework from the perspective of different network functionalities. 
Placing content at the network edge has been shown to reduce the end-to-end latency and degrade the increased traffic on the origin server~\cite{Wang2014,Ge2016}. In computationally intensive video processing applications such as augmented reality (AR) or virtual reality (VR), computing tasks of mobile clients can be executed at the edge servers, which reduces the computational burden of a central cloud. The potential of edge caching in reducing duplicate access of clients to popular content located on the origin server has also been demonstrated. \emph{Wang et. al} \cite{Wang2014} analyze the impact of content caching at different layers of the mobile network. 


From the applications point of view, video streaming users are the most bandwidth hungry. Their bandwidth demand is expected to dramatically increase with the next generation mobile networks~\cite{Taleb2017,Tran2017}. With on-demand streaming, serving popular video segments with large media size from the edge of the network reduces congestion on the backhaul of the network, hence, providing more bandwidth for other services \cite{Taleb2017}. Depending on the architecture of the edge network, the quality of experience (QoE) of mobile users can be improved by reducing the non-negligible delay with origin server through edge caching. \emph{Ge et. al} \cite{Ge2016} have analyzed the achieved gain in QoE by caching the video segments with multiple qualities at the edge server. However, their model is applicable for the services with non-negligible delay between the clients and the origin video server. Furthermore, there is no concrete optimization framework proposed in this work for jointly maximizing the QoE of the clients and reducing the clients access to the origin content server.   


In the context of dynamic adaptive video streaming over HTTP (DASH), several approaches for improving QoE have been proposed during the past years~\cite{Huang2014,Mangla2016,Wang2016}. \emph{Seufert et. al} \cite{Seufert2015} provides a comprehensive study on DASH quality adaptation and the major factors that both client and network have to take into account. On the client side, the quality adaptation approaches adjust the requested video bitrate according to network conditions by relying on either the instantaneous buffer occupancy level~\cite{Huang2014,Spiteri2016} or estimated bandwidth according to the previously perceived throughput \cite{Mangla2016}, or a combination of the two mechanisms~\cite{Wang2016}. 
Some research also investigated the scalability of DASH strategies~\cite{Chen2013,Petrangeli2015,Bethanabhotla2015,Bouten2014} when multiple clients are associated with either single or multiple video servers. 
Although these papers consider large set of mobile clients, relying on the client-based adaptation heuristics \cite{Chen2013,Petrangeli2015}, the bitrates may not be fairly allocated to the clients due to the lack of coordination among them in some situations such as the interleaving of their arrival and departure times. This in turn causes the unfairness among the competing clients which are not handled by the client-based adaptation strategies. Furthermore, the focus of the work in \cite{Chen2013} is mainly on the fair resource allocation than the QoE of the clients since the major QoE factors have not been involved in the optimization problem. Similarly, some important factors such as stalling event which has significant impact on the perceived QoE \cite{Seufert2015}, has not been taken into account in the in-network adaptive optimization framework proposed in \cite{Bouten2014}.      


Server and network assisted DASH (SAND-DASH) standard has been recently published. It provides mechanisms for collaboration between the mobile clients and in-network elements. The first research efforts presented in ~\cite{Li2016,Cofano2016,Thomas2016} investigate the effectiveness of SAND-DASH standard through experimental setup. 
Compared to our work, they do not provide a concrete optimization framework for the joint maximization of QoE of the clients and the fair resource allocation at the network side. 
Authors in \cite{Mehrabi2017} design an optimization problem for jointly maximizing the QoE, the fair bitrate allocation among the competing clients as well as balancing the utilized resources among multiple edge servers. However, they do not exploit the potential of content caching at the edge servers which results in significant savings in the operational costs for ISPs. QoE-driven content caching at either the edge server or cellular base station has been also suggested in~\cite{Zhang2013,Ahlehagh2014}. However, the scalability/fairness issue and the imperfectness of the factors involved in QoE objectives are the main limitations of these works. 


Joint video caching and processing at the edge of the network has also been studied. 
\emph{Pedersen et. al} \cite{Pedersen2016} propose the joint optimization of adaptive video streaming, backhaul resource allocation, and video content caching at RAN. The proposed adaptive video streaming and caching model in this work assumes that the clients request to the chunks with lower bitrates may be served from the cache by transcoding from the available chunks with higher bitrates at the edges. 
To efficiently utilize the potential of joint adaptive video streaming and caching, they also propose a proactive user preference profile (UPP)-based cache replacement strategy. 
However, in scenarios with limited processing capabilities at the edge servers and high bandwidth capacity on the backhaul network, the encoding of the video chunks can be performed at the origin server while the caching and clients information processing are handled at the edges. Furthermore, the proposed cache replacement strategy in \cite{Pedersen2016} takes into account only the video popularity and bitrate estimation, while, some statistical information about the clients departure (retention) toward different videos is another factor that can be utilized to improve the LRU-based cache replacement strategies. 


With low access delay to the cloud server using fast-ultra communication lines in the next generations of mobile networks \cite{Chen2016}, it is essential for mobile network operator (MNO) to decide on a desired trade-off between the created backhaul traffic and the QoE of the mobile clients in joint adaptive video streaming and edge caching. Under such scenario, the impact of video caching on the QoE of the clients (i.e. the average video bitrate, the switching frequency and magnitude) has not been well exploited. Furthermore, none of the current studies investigate how quantitatively the periodic cache updating impacts the QoE of the clients with respect to the traffic reduction on the backhaul network. To address these impacts, we propose the joint QoE-traffic optimization for edge caching in MEC environments. The proposed optimization problem and its variations provide a setup to analyze the impact of caching and content replacement on the QoE of the clients under different scenarios. 
We further propose an effective retention-based cache updating heuristic which is handled at the edges independent from the optimization problem. Inspired by the research work in \cite{siekkinen16ton}, we integrate the retention of the clients during different parts of the video into our proposed cache replacement heuristic which as we show through simulations results in significant reduction in cache miss rate compared to LRU-based policies.   

\begin{figure}[t]
\begin{center}
\includegraphics[trim = 0in 0in 0in 0in, clip=true, width=3.4in, height=2.3in]{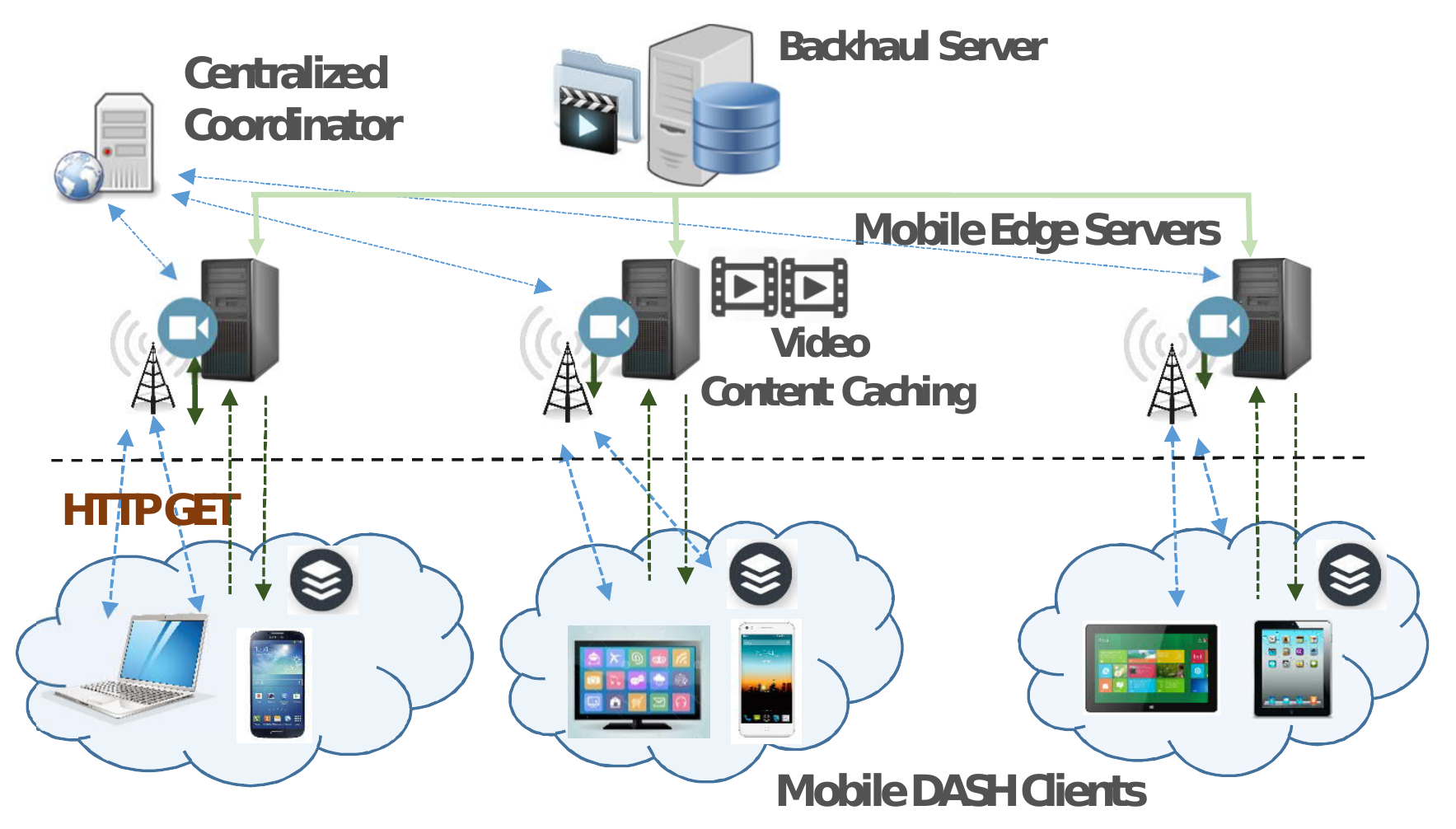} 
\caption{Mobile edge caching adaptive video streaming.} 
\label{fig:framework}
\vspace{-2em}
\end{center}
\end{figure}

\section{Mobile Edge Assisted Adaptive Streaming with Caching} 
\label{sec:framework}

\subsection{System Overview}
\label{subsec:overview}

Fig. \ref{fig:framework} illustrates our target system for edge assisted adaptive video streaming with content caching. As with any HTTP based streaming, video content is divided into chunks of fixed size $C$ (in seconds) and each chunk is stored with different discrete qualities (bitrates) denoted by set $R$ on the origin server in the cloud. The clients' requests and video caching are handled by the edge servers located within the radio access network (RAN). Edge servers are associated with base stations (eNodeBs) which allocate the available radio resource blocks to the local clients in a proportionally fair manner~\cite{Chen2013}.

The (centralized) coordinator receives video stream level information from mobile client applications and radio link information from the eNodeBs, both via the edge servers. Using this information, the coordinator periodically solves an optimization problem (Section \ref{sec:optimizationproblem}) and forwards the solutions to the edge servers which guide the clients in bitrate selection process. The solution balances two opposing optimization targets: minimization of video traffic flowing through the backhaul network (and cross the ISP if the next closest video server that we call origin video server lies outside of the ISP) and maximization of client perceived QoE. Striving only for the former target would make the solution always favor already cached video chunks even if their bitrate is smaller than the bitrate that the available radio access resources would allow downloading without the risk of buffer underrun. In contrast, striving only for the latter target would make the solution ignore the cache status and always download as high bitrate chunk as is possible given the capacity constraint of the network. 

We should note that multiple coordinators can be considered for different groups of the edge servers in our system design which in turn facilitates the decentralized implementation of joint DASH video streaming and edge caching. It is also important to note that our system design does not modify the operation of the eNodeBs in any way. We merely require the existence of edge computing infrastructure with edge servers that are able to receive radio link level information from the eNodeBs and mobile video streaming applications that cooperate with the edge servers. 


\subsection{System Notations}

We consider the discrete time slotted DASH scheduling \cite{Bouten2014} with total number of $|T|$ time slots of $\Delta t$ seconds each. At every time slot $1 \leq t \leq |T|$, video content from edge server $k$ is transmitted by the eNodeB associated with over the shared radio access having capacity of $W_{k}^{(t)}$. Please note that $W_{k}^{(t)}$ refers to the available resource blocks i.e. the number of subcarriers in frequency domain, at base station $k$ in slot $t$. 

Let $A_{i}$ and $D_{i}$ denote the arrival and departure time slots, respectively, of client $i$ which correspond to the time that client sends its request for first chunk and the time that it either abandons the streaming session or finishes downloading the last chunk. In the ideal case when no stalling happens during the session and with negligible network delay, the quantity $(D_{i}-A_{i})$ is equal to the watching duration of the video by client $i$ and consequently $\lceil (D_{i}-A_{i})/C \rceil$ is the number of chunks streamed. Notation $v_{i}$ represents the video index which is watched by client $i$. 


The media player of each client $i$ maintains a playback buffer with maximum capacity (in $Mb$) denoted by $B_{i}^{max}$. $0 < B_{i}^{(t)} \leq B_{i}^{max}$ represents the level of data in the client's buffer at time slot $t$. The parameter $SNR_{ik}^{(t)}$ is also defined which denotes the received signal-to-noise ratio (SNR) by client $i$ from base station $k$ at time slot $t$. For client to edge server mapping, we define the binary indicator $a_{ik}^{(t)}$ such that $a_{ik}^{(t)}=1$ when the client $i$ is allocated to server $k$ for downloading the current chunk at time slot $t$ and $a_{ik}^{(t)}=0$, otherwise. In our system model, we assume that the server allocation is decided at the beginning of each new chunk and the current server remains unchanged if the client is downloading at the middle of the chunk. At the beginning of each new chunk, the client is mapped to the server from where it receives the highest SNR value from the local BS at the corresponding time slot. Also, the integer decision variable $r_{ik}^{(t)} \in R$ denotes the allocated bitrate for chunk index $p$ of client $i$ at server $k$. At each time slot $t$, the request of each client $i$ is represented by triple $(v_{i},\lceil (t-A_{i})/C \rceil,r_{ik}^{(t)})$ where the first item refers to the requested video while the second and third items denote the index and bitrate of the requested chunk. Furthermore, the notation $M_{k}^{(t)}$ represents the set of above-mentioned triples which are available in the cache of edge server $k$ at time slot $t$. Binary decision variable $d_{ik}^{(t)}$ is also defined such that $d_{ik}^{(t)}=1$ indicates that at time slot $t$, client $i$ at server $k$ downloads its chunk/bitrate directly from the origin server and $d_{ik}^{(t)}=0$ if its request exists in the cache of the local edge server. All edge servers have a cache of equal size $Q$. For the sake of simplicity, we have listed the set of parameters involved in the system model together with their descriptions in Table \ref{table:systemparameters}. We discuss next the different optimization criteria related to QoE, fairness, and data traffic.

\begin{table}[t]
\renewcommand{\arraystretch}{1.3}
\caption{System notations and their descriptions.}
\label{table1}
\centering
\begin{scriptsize}
\begin{tabular}{|p{2cm}|l|}
\hline
\textbf{Notation} & \multicolumn{1}{|c|}{\textbf{Description}}  \\
\hline\hline
$C$ & \multicolumn{1}{p{6cm}|}{Constant size of each video chunk (in seconds)}  \\
\hline 
$K, S, R$   &  \multicolumn{1}{p{6cm}|}{Number of edge servers, clients and the discrete set of available video bitrates, respectively}    \\
\hline
$|T|$, $\Delta t$   & \multicolumn{1}{p{6cm}|}{Total number of scheduling time slots and the duration of each slot in seconds} \\
\hline
$W_{k}^{(t)}$, $M_{k}^{(t)}$    &   \multicolumn{1}{p{6cm}|}{Available resource blocks at base station $k$ and the set of cached chunks at edge server $k$ in time slot $t$, respectively}    \\
\hline
$Q$   & \multicolumn{1}{p{6cm}|}{Constant cache size} \\
\hline
\hline
$A_{i}, D_{i}$   & \multicolumn{1}{p{6cm}|}{Arrival and departure times of client $i$} \\
\hline
\hline
$B_{i}^{max}$   &  \multicolumn{1}{p{6cm}|}{Maximum buffer capacity (in $Mb$) for client $i$}  \\
\hline
$B_{i}^{(t)}$   &  \multicolumn{1}{p{6cm}|}{Buffer level of client $i$ at time slot $t$}  \\
\hline
$AQ_{i}$   &  \multicolumn{1}{p{6cm}|}{Average video quality for client $i$}  \\
\hline
$L_{i}$   &  \multicolumn{1}{p{6cm}|}{Initial delay on the client $i$'s buffer}  \\
\hline
$E_{i}$    &  \multicolumn{1}{p{6cm}|}{Accumulated bitrate switching for client $i$} \\
\hline
$SNR_{ik}^{(t)}$, $Thr_{ik}^{(t)}$, $\hat{Thr}_{ik}^{(t)}$   &  \multicolumn{1}{p{6cm}|}{Received SNR, theoretical data throughput and effective throughput by client $i$ from base station $k$ at time slot $t$}  \\
\hline
\hline
$F_{i}$   &  \multicolumn{1}{p{6cm}|}{Fairness of client $i$} \\
\hline
$BT_{i}$  &  \multicolumn{1}{p{6cm}|}{Overall downloaded data by client $i$ from the backhaul origin server}   \\
\hline
\hline
$\rho$, $\omega$, $\gamma$   &  \multicolumn{1}{p{6cm}|}{Adjustable weighting parameters for average quality, bitrate switching and fairness respectively}  \\
\hline
$\beta$   &  \multicolumn{1}{p{6cm}|}{Weighting coefficient for controlling the importance of QoE and data traffic in the joint optimization}   \\
\hline
\hline
$a_{ik}^{(t)}$   &   \multicolumn{1}{p{6cm}|}{Binary indicator for the allocation of client $i$ to server $k$ at time slot $t$}  \\
\hline
$d_{ik}^{(t)}$   &   \multicolumn{1}{p{6cm}|}{Binary indication that the bitrate of the current chunk of client $i$ at server $k$ in time slot $t$ is downloaded from the origin server}  \\  
\hline
$r_{ik}^{(p)} \in R$    &   \multicolumn{1}{p{6cm}|}{Discrete allocated bitrate to chunk index $p$ of client $i$ at server $k$}   \\   
\hline
$(v_{i},\lceil (t-A_{i})/C \rceil,r_{ik}^{(t)})$   &  \multicolumn{1}{p{6cm}|}{Requested $(video,chunk,bitrate)$ by client $i$ at time slot $t$}   \\ 
\hline    
\end{tabular}
\end{scriptsize}
\label{table:systemparameters}
\end{table}

\subsection{Quality of Experience}
\label{subsec:quality}

Studies show that following four factors significantly affect the quality of experience perceived by adaptive video streaming clients: \emph{video quality}, \emph{startup delay}, \emph{stalling ratio} and \emph{bitrate switching}~\cite{Seufert2015}. 

\textbf{Video bitrate} 
Video bitrate is directly related to the perceived quality of the video. Heavier compression (quantization) produces smaller bitrate but also worse perceptual quality. 
There is also a trade-off between video bitrate and stalling: High bitrate streaming increases the probability of experiencing a stall event because the download throughput has a higher chance to drop below that bitrate due to contention at the bottleneck link or reduced wireless link quality. 
The average video bitrate over $(D_{i}-A_{i})/C$ downloaded chunks by client $i$ is obtained using the following relation: 
\begin{align}
\label{eq1}
AQ_{i}=\lceil \frac{C}{D_{i}-A_{i}} \rceil \sum_{p=1}^{\lceil (D_{i}-A_{i})/C \rceil} \sum_{k=1}^{K} a_{ik}^{(A_{i}+(p-1) \cdot C)} \cdot r_{ik}^{(p)}
\end{align}

\textbf{Startup delay} refers to the time duration needed to fill up the playback buffer upon the arrival of a client. Hence, it is the waiting time of client from click to start of the playback. According to~\cite{Hossfeld2012}, the startup delay has a clearly smaller impact on the dissatisfaction of a viewer than stall events. Denoted by $L_{i}$ as the time delay to reach the maximum buffer filling level $B_{i}^{max}$ for client $i$, the following equality must be satisfied:  
\begin{align}
\label{eq2}
&\sum_{t=A_{i}}^{A_{i}+L_{i}} \sum_{k=1}^{K} a_{ik}^{(t)} \cdot \hat{Thr}_{ik}^{(t)} \cdot \Delta t=B_{i}^{max},
\end{align} 
where $\hat{Thr}_{ik}^{(t)}$ is the effective data throughput (in $Mbps$) received by client $i$ from server $k$ at time slot $t$. For computing the theoretical downlink throughput $Thr_{ik}^{(t)}$ over the wireless link, we employ the following Shannon upper bound approximation: 
\begin{align}
\label{eq3}
\small
Thr_{ik}^{(t)}=
\begin{cases}
0   \hspace{15mm}  \text{$SNR_{ik}^{(t)} < SNR_{min}$}   \\
\alpha \cdot log_{2}(1+10^{\frac{SNR_{ik}^{(t)}}{10}})    \\
   \hspace{17mm}    \text{$SNR_{min} \leq SNR_{ik}^{(t)} < SNR_{max}$}   \\
Thr_{max}   \hspace{15mm}    \text{$SNR_{ik}^{(t)} \geq SNR_{max}$}  
\end{cases}
\end{align}
Where coefficient $\alpha$ and parameters $SNR_{min}$, $SNR_{max}$ and $Thr_{max}$ are set to 0.6, -10 dB, 23 dB and 4.4 bps/Hz respectively, according to the LTE downlink specifications reported in \cite{LTEThroughput}. 
Note that the effective share throughput of client $i$ is computed by the relation $\hat{Thr}_{ik}^{(t)}=(Thr_{ik}^{(t) 2}/\sum_{\forall j} a_{jk}^{(t)} \cdot Thr_{jk}^{(t)}) \cdot W_{k}^{(t)}$ where the summation in denominator is taken over all clients $j$ which have been assigned to base station $k$ at time slot $t$.    

\textbf{Stalling ratio} is the the amount of time spent so that video playback is stalled divided by the total duration of the session. Stall events occur when playback buffer empties caused by too low download throughput compared to the video bitrate. Avoiding stall events is critically important because of their prominent role in determining QoE. As for avoiding stall events, we assume that the player starts to play the video after the startup phase. Given $\hat{Thr}_{ik}^{(t)}$, the buffer level (in $Mb$) of client $i$ at time slot $t$ is given by:  
\begin{align}
\label{eq4}
B_{i}^{(t)}=
\begin{cases}
 B_{i}^{(t-1)}+\hat{Thr}_{ik}^{(t)} \cdot \Delta t, \hspace{7mm} A_{i} \leq t \leq A_{i}+L_{i}  \\
 B_{i}^{(t-1)}+(\hat{Thr}_{ik}^{(t)}-r_{ik}^{(p)}) \cdot \Delta t, \\
 \hspace{41mm}  A_{i}+L_{i} < t \leq D_{i}   
 \end{cases}
\end{align}
where $r_{ik}^{(p)}$ is the allocated bitrate for the currently played out chunk with index $p$. Accounting for the arrival time of client and initial playback delay, the index $p$ of the chunk played out at time slot $t > A_{i}+L_{i}$ is equal to $p=\lceil (t-A_{i}-L_{i})/C \rceil$. Later, we design the optimization problem with such constraints that stall events are avoided whenever possible, i.e. whenever the total amount of resources suffices to support lowest available video bitrates for all clients. 

Frequent \textbf{bitrate switching} is also considered harmful for QoE~\cite{Seufert2015}. We consider the difference between the bitrates of consecutive chunks of the video downloaded by the client as the QoE metric for switching. Hence, the accumulated bitrate switching for client $i$ during the streaming session is given by: 
\begin{align}
\label{eq5}
E_{i}=\sum_{p=2}^{\lceil (D_{i}-A_{i})/C \rceil} \sum_{k=1}^{K} \{&a_{ik}^{(A_{i}+(p-1) \cdot C)} \cdot r_{ik}^{(p)} \nonumber  \\
&-a_{ik}^{(A_{i}+(p-2) \cdot C)} \cdot r_{ik}^{(p-1)}\}
\end{align}
We should note that the proposed framework in this work is easily adoptable to different relations for the switching without any change in the analytical model.

\subsection{Fairness}
\label{subsec:fairness}

LTE base stations usually allocate the radio resource blocks to the mobile clients in a proportional fair (PF) manner according to their wireless link quality ~\cite{Chen2013}. 
Based on the share of the allocated resources by the base station, each mobile DASH client then chooses the most sustainable bitrate. However, due to the lack of coordination among multiple DASH clients sharing the radio access link, the adaptation heuristics on the client side may allocate the bitrates in an unfair manner in some situations \cite{Chen2013}.  

As a part of our system design, we incorporate the fairness in bitrate allocation such that at each time slot, the best sustainable bitrate is selected for the client which has the least difference from the average of bitrates allocated to the other simultaneous clients at the same time slot. More precisely, the objective of fair bitrate allocation is to minimize the overall bitrate deviations of each client $i$ during its whole video streaming session which is obtained using the following relation: 
\begin{align}
\label{eq6}
F_{i}=\sum_{t=A_{i}}^{D_{i}} \sum_{k=1}^{K} a_{ik}^{(t)} \cdot (r_{ik}^{(t)}-\bar{r}^{(t)})  
\end{align}
Where $\bar{r}^{(t)}=(1/N^{(t)})\sum_{\forall j \neq i}\sum_{1 \leq p \leq K} a_{jp}^{(t)} \cdot r_{jp}^{(t)}$ is the average bitrates of other $N^{(t)}=\sum_{\forall j \neq i}\sum_{1 \leq p \leq K} a_{jp}^{(t)}$ simultaneous active clients at time slot $t$. It should be noted that for each individual client, the minimization of $F_{i}$ should satisfy the available resource blocks at the base station in each time slot.

\subsection{Data Traffic}
\label{subsec:datadownload}

In case that the requested chunk/bitrate of the clients's streaming video does not exist in the edge server's cache, the client downloads the video chunk and bitrate from the origin server which in turn generates data traffic on the backhaul network, and possibly cross ISP depending on the location of the origin video server. In our system model, we consider the volume of video data downloaded from the origin server at each time slot as the metric used in the joint QoE-traffic optimization problem. 

Considering the binary decision variable $d_{ik}^{(t)}$ as the indication of downloading the chunk from the origin server, the overall volume of data traffic from origin server caused by client $i$ during its whole video streaming session, denoted by $BT_{i}$, is obtained using the following summation:
\begin{align}
\label{eq7}
BT_{i}=\sum_{t=A_{i}}^{D_{i}} \sum_{k=1}^{K} a_{ik}^{(t)} \cdot d_{ik}^{(t)} \cdot r_{ik}^{(t)}
\end{align}

\section{QoE-traffic Optimization Problem} 
\label{sec:optimizationproblem}

We formulate the joint optimization of QoE and video traffic in this section. In accordance with Section \ref{subsec:quality}, we define three adjustable weighting parameters $0 \leq \rho, \omega, \gamma \leq 1$ to control video quality, bitrate switching, and fairness, respectively. In addition, the weight $0 \leq \beta \leq 1$ controls the importance of QoE-fairness and the volume of video data downloaded from the origin server, respectively. Furthermore, for each individual client, we include a constraint in the optimization problem in order to avoid playback stall events during the whole streaming process. 


With the parameters defined in section \ref{sec:framework} and relations (\ref{eq1}),(\ref{eq2}),(\ref{eq4}),(\ref{eq5}),(\ref{eq6}) and (\ref{eq7}) the joint optimization for each client $i$ is defined as a utility maximization problem with the following integer non-linear programming (INLP) formulation: 
\begin{align}
\label{obj}
\underset{d,r}{Maximize} \hspace{3mm} U_{i}= \beta(\rho AQ_{i} - \omega E_{i} - \gamma F_{i}) - (1-\beta) BT_{i}
\end{align}
Subject to:
\begin{align}
\label{cons1}
&\sum_{j \in S} a_{jk}^{(t)} \cdot \lceil \frac{r_{jk}^{(\lceil t/C \rceil)}}{Thr_{jk}^{(t)}} \rceil  \leq W_{k}^{(t)},    \\
&  \hspace{25mm} \forall 1 \leq k \leq K, \hspace{2mm} 1 \leq t \leq |T|  \nonumber  \\
\label{cons2}
& d_{ik}^{(t)}=\mathbb{I}((v_{i},\lceil (t-A_{i})/C \rceil,r_{ik}^{(t)}) \notin M_{k}^{(t)}),     \\
& \hspace{25mm} 1 \leq k \leq K, \hspace{2mm} A_{i} \leq t \leq D_{i}   \nonumber   \\ 
\label{cons3}
& 0  < B_{i}^{(t)}  \leq B_{i}^{max}, \hspace{5mm} \forall A_{i} \leq t \leq D_{i}  \\ 
\label{cons4}
& a_{ik}^{(t)}=    
\begin{cases}
a_{ik}^{(t-1)},  \hspace{3mm} \text{$t$ $mod$ $C$ $\neq$ $1$}  \\  
1,  \hspace{3mm} \text{$t$ $mod$ $C$ $=$ $1$ $\wedge$ $k=arg \hspace{1mm} \textbf{max}\{SNR_{ik}^{(t)}\}$}  \\   
0, \hspace{3mm} \text{Otherwise}      
\end{cases}  \\ 
\label{cons5}
& r_{ik}^{(p)} \in R, \hspace{2mm} d_{ik}^{(t)} \in \{0,1\},   \\ 
&  \hspace{5mm}  \forall 1 \leq k \leq K, A_{i} \leq t \leq D_{i}, 1 \leq p \leq (D_{i}-A_{i})/C   \nonumber 
\end{align} 
Note that in the above optimization model, the equality (\ref{eq2}) is also added to the set of constraints. The only decision variables are $d_{ik}^{(t)}$ and $r_{ik}^{(p)}$ which are respectively the binary and integer variables and the values of other parameters are known in advance. 
The objective function (\ref{obj}) aims to maximize jointly the QoE of DASH client $i$, the fairness as well as minimizing the overall downloaded video data from the backhaul sever by the client. Constraint (\ref{cons1}) ensures that at any time slot and for any base station, the available resource blocks are assigned to the clients proportional to their throughput (link quality). Constraint (\ref{cons2}) states that at each time slot the current chunk of video is downloaded from the origin server if the chunk is not available in the cache of the associated edge server. Note that the identity function $\mathbb{I}((v_{i},\lceil (t-A_{i})/C \rceil,r_{ik}^{(t)}) \notin M_{k}^{(t)})=1$ if $(v_{i},\lceil (t-A_{i})/C \rceil,r_{ik}^{(t)}) \notin M_{k}^{(t)}$ and $\mathbb{I}((v_{i},\lceil (t-A_{i})/C \rceil,r_{ik}^{(t)}) \notin M_{k}^{(t)})=0$ if $(v_{i},\lceil (t-A_{i})/C \rceil,r_{ik}^{(t)}) \in M_{k}^{(t)}$. 
Constraint \ref{cons3} ensures that no stalling happens on the client's buffer and the relation (\ref{cons4}) determines the edge server to which the client is mapped at each time slot. 
It should be noted that $a_{ij}$'s values are determined beforehand and are not counted as decision variables for the optimization problem. Finally, the set of constraints in (\ref{cons5}) specify the range of the decision variables.  


\section{Online Algorithms}
\label{sec:algorithm}

The joint optimization problem formulated in (\ref{obj})-(\ref{cons5}) belongs to the class of NP-hard problems due to the existence of integer decision variables in the model and hence the exhaustive possible enumerations for the solution space. When all the information of clients (the arrival and departure time slots, SNR values) are known in advance, the straightforward brute-force search strategy can be employed to investigate all possibilities of allocating bitrates to clients and select the one with maximum achievable utility. However, the computational complexity of brute-force strategy significantly grows with the increase in the number of clients or the set of available bitrates making it infeasible for MNOs to deploy the solution in large scale scenarios. To reduce the complexity, we design a greedy-based algorithm with low computational complexity for online operation. The pseudo-code of the proposed algorithm named as cache-based self-tuned bitrate allocation algorithm (CSBAA) has been illustrated in Algorithm \ref{algorithm}. 

\begin{algorithm}[t]
\small
\caption{Cache-based Self-tuned Bitrate Allocation Algorithm (CSBAA) (Run by the centralized coordinator)}
\label{algorithm}
\begin{algorithmic}[1]
\small 
\State \parbox[t]{\dimexpr\linewidth-\algorithmicindent}{\textbf{Input:} \hspace{0.2mm} $|T|, K, R:$ Number of scheduling time slots, number of DASH edge servers, set of available discrete bitrates on the origin server. \strut} 
\State \parbox[t]{\dimexpr\linewidth-\algorithmicindent}{\textbf{Output:} \hspace{0.2mm} Binary allocation $d_{ik}^{(t)}$ and integer bitrate allocation $r_{ik}^{(t)}$ for each client $i$, edge server $1 \leq k \leq K$ and time slot $1 \leq t \leq |T|$, $totalUtility$\strut}
\vspace{0.1em} 
\State \parbox[t]{\dimexpr\linewidth-\algorithmicindent} {$M_{k}^{(t)}=\emptyset \hspace{3mm} \forall 1 \leq k \leq K, \forall 1 \leq t \leq |T|$; \strut} 
\For{each time slot $1 \leq t \leq |T|$}
   \For{each client $i$ such that $A_{i} \leq t \leq D_{i}$}
      \State \parbox[t]{\dimexpr\linewidth-\algorithmicindent}{\textbf{Allocate} client $i$ to server $1 \leq k \leq K$  \\ 
      \indent \hspace{1mm} according to (\ref{cons4}) \strut}    
      \If{$t = A_{i}$}
        \State \parbox[t]{\dimexpr\linewidth-\algorithmicindent} {\textbf{Initialize} $BufferStatus$, $BT_{i}$ and $L_{i}$ \strut}  
      \EndIf
      \If{$(t-A_{i})$ \textbf{mod} $C \neq 1$}
          \State \parbox[t]{\dimexpr\linewidth-\algorithmicindent}{\textbf{Allocate} client $i$ to the same server and \\ 
          with the same bitrate as with time slot $t-1$; \strut}  
            \State {\textbf{Update} $B_{i}^{(t)}$, $BT_{i}$;}            
            \State \parbox[t]{\dimexpr\linewidth-\algorithmicindent}{\textbf{if} $BufferStatus = False$ \textbf{And} \\
            \indent \hspace{12mm} $B_{i}^{(t)} = B_{i}^{max}$ \textbf{then} \strut}
            \State {\indent \hspace{3mm} $BuffetrStatus=True; L_{i}=t-A_{i}$;}
      \EndIf
      \If{$(t-A_{i})$ mod $C = 1$}
         \If{$BufferStatus = False$}
             \State \parbox[t]{\dimexpr\linewidth-\algorithmicindent}{\textbf{Call} Subroutine Startup Phase; \strut}
         \EndIf
         \If{$BufferStatus = True$}
             \State \parbox[t]{\dimexpr\linewidth-\algorithmicindent}{\textbf{Call} Subroutine Steady State; \strut}
         \EndIf 
      \EndIf
  \If{$t = D_{i}$}
      \State  \parbox[t]{\dimexpr\linewidth-\algorithmicindent}{$totalUtility=totalUtility+U_{i}$ \strut}   
   \EndIf   
  \EndFor
  \For{each edge server $1 \leq k \leq K$}
  \If{$d_{ik}^{(t)} = 1$, \hspace{1mm} for at least one client $i: a_{ik}^{(t)}=1$} 
   \State \parbox[t]{\dimexpr\linewidth-\algorithmicindent}{\textbf{Call} Subroutine RBCRH for server $k$ \\
    \indent \hspace{1mm} at time slot $t$;  \strut}
  \EndIf
  \EndFor
 \EndFor
\State  \parbox[t]{\dimexpr\linewidth-\algorithmicindent}{\textbf{Return} $totalUtility$; \strut}
\end{algorithmic}
\end{algorithm}

\subsection{Self-tuned Bitrate Selection}
\label{subsec:selftuned}

The subroutine Startup Phase is called at each time slot for all the clients whose playback buffer has not yet been filled up. Otherwise, the client is in the steady state and the Subroutine 2 is invoked. We assume that the mobile network always assigns each client to the (nearest) eNodeB providing the highest throughput. If the client is in the middle of downloading a chunk, it merely continues. Otherwise, the algorithm first chooses such a sustainable bitrate to the client that results in least amount of switching and highest fairness. More precisely, the scheduler first considers a bitrate whose difference to the bitrate of the previous chunk is less than some threshold $\delta_{S}$ and that results in a fairness value greater than or equal to the threshold $\delta_{F}$. Based on the result from our work in \mbox{\cite{Mehrabi2017}}, the highest amount of bitrate switching happens when the client-based heuristic allocates the bitrates merely based on the playout buffer occupancy level.
Therefore, we compute the amount of switching when the buffer-based adaptation strategy is applied for bitrate allocation and use that amount as the switching threshold $\delta_S$.   
In fact, our algorithm allocates a bitrate to the current chunk of the client which its resulting switching is less than the threshold $\delta_S$. 
The fairness threshold $\delta_F$ is also provided to the algorithm at the deployment phase. 

The decision on bitrate selection is made considering two possibilities: Either the current chunk/bitrate of the streaming video exists in the cache of the associated edge server or it has to be downloaded from the origin server. The algorithm makes the decision on which bitrate to allocate by taking into account the QoE-traffic trade-off according to the weighting parameter $\beta$ in objective function (\ref{obj}). After the bitrate was allocated to the client, the weighting values in the QoE term of the optimization problem are dynamically adjusted. The weighting of $AQ$ term is determined based on how far is the selected bitrate from the maximum available one, As for the switching term $E$, its weight is derived based on how far the chosen bitrate is from the one that yields no switching (same as previous chunk). The weight of the fairness term $F$ is similarly computed according to the difference between the selected and the average bitrates of other clients.  \\ \\ 
\begin{tabular}{p{7.6 cm}}
\hline 
\textbf{Subroutine 1:} Startup Phase \\
\hline \\
\end{tabular}
\begin{algorithmic}[1]
\small 
\vspace{-1em}
           \If{$t-A_{i} \leq C$}
                \State{\textbf{Allocate} the highest available bitrate;}
                \State{\textbf{Update} \emph{BufferStatus}, $B_{i}^{t}$ and $BT_{i}$ \strut} 
                \EndIf
           \State {\textbf{Compute} $estThr$ from the allocated server;} 
      \State {\textbf{Compute} switching thresholds $\delta_{S}$;}  
      \State {$maxUtility=- \infty$;} 
        \For{each bitrate $r \in R$ in decreasing order}
          \State \parbox[t]{\dimexpr\linewidth-\algorithmicindent}{\textbf{if} allocation of $r$ satisfy (\ref{cons1}) \textbf{AND}  \\
          \indent \hspace{4mm} $r \leq \textbf{max}(estThr,\hat{Thr}_{ik}^{(t)},B_{i}^{(t)})$ \textbf{then} \strut}
             \State \parbox[t]{\dimexpr\linewidth-\algorithmicindent}{\indent  \hspace{6mm} \textbf{if} $|r-r_{ik}^{(t-1)}| \leq \delta_{S}$ \textbf{AND}  \\  
           \indent  \hspace{10mm} $1-|r-\bar{r}|/(R_{max}-R_{min})>=\delta_{F}$ \textbf{then} \strut}   
           \State {\indent \hspace{10mm} $Data=0;$}  
          \State {\indent \hspace{10mm} \textbf{if} $(v_{i},\lceil \frac{t-A_{i}}{C} \rceil,r) \notin M_{k}^{(t)}$ \textbf{then} $Data=r$;}  
           \State {\indent \hspace{10mm} \textbf{Compute} weightings $\rho,\omega$ and $\gamma$;}  
           \State {\indent \hspace{10mm} $QE=\rho r - \omega |r-r_{ik}^{(t-1)}| - \gamma |r-\bar{r}|$;}  
           \State {\indent \hspace{10mm} \textbf{if} $\beta QE - (1-\beta) Data > maxUtility$ \textbf{then}}  
                \State {\indent \hspace{14mm} $maxUtility=\beta QE - (1-\beta) Data$;}  
              \State {\indent \hspace{14mm} $r_{ik}^{(t)}=r$;}   
         \EndFor
         \If{$r_{ik}^{(t)} = 0$}
             \For{each bitrate $r \in R$ in decreasing order}
             \State \parbox[t]{\dimexpr\linewidth-\algorithmicindent}{\textbf{if} allocation of $r$ satisfy (\ref{cons1}) \textbf{AND}  \\
          \indent \hspace{4mm} $r \leq \textbf{max}(estThr,\hat{Thr}_{ik}^{(t)},B_{i}^{(t)})$ \textbf{then} \strut} 
             \State \parbox[t]{\dimexpr\linewidth-\algorithmicindent}{\hspace{6mm} \textbf{if} $|r-r_{ik}^{(t-1)}| \leq \delta_{S}$ \textbf{then} \strut} 
            \State {\indent \hspace{5mm} $Data=0;$}   
          \State {\indent \hspace{5mm} \textbf{if} $(v_{i},\lceil \frac{t-A_{i}}{C} \rceil,r) \notin M_{k}^{(t)}$ \textbf{then} $Data=r$;} 
           \State {\indent \hspace{5mm} \textbf{Compute} weightings $\rho,\omega$ and $\gamma$;} 
           \State {\indent \hspace{5mm} $QE=\rho r - \omega |r-r_{ik}^{(t-1)}| - \gamma |r-\bar{r}|$;} 
           \State {\indent \hspace{5mm} \textbf{if} $\beta QE - (1-\beta) Data > maxUtility$ \textbf{then}} 
               \State {\indent \hspace{9mm} $maxUtility=\beta QE - (1-\beta) Data$;} 
            \State {\indent \hspace{9mm} $r_{ik}^{(t)}=r$;} 
            \EndFor
         \EndIf
         \If{$r_{ik}^{(t)} = 0$}
             \For{each bitrate $r \in R$ in decreasing order}
             \State \parbox[t]{\dimexpr\linewidth-\algorithmicindent}{\textbf{if} allocation of $r$ satisfy (\ref{cons1}) \textbf{AND}  \\
          \indent \hspace{4mm} $r \leq \textbf{max}(estThr,\hat{Thr}_{ik}^{(t)},B_{i}^{(t)})$ \textbf{then} \strut} 
            \State {\indent \hspace{5mm} $Data=0;$} 
          \State {\indent \hspace{5mm} \textbf{if} $(v_{i},\lceil \frac{t-A_{i}}{C} \rceil,r) \notin M_{k}^{(t)}$ \textbf{then} $Data=r$;} 
           \State {\indent \hspace{5mm} \textbf{Compute} weightings $\rho,\omega$ and $\gamma$;} 
           \State {\indent \hspace{5mm} $QE=\rho r - \omega |r-r_{ik}^{(t-1)}| - \gamma |r-\bar{r}|$;}
           \State {\indent \hspace{5mm} \textbf{if} $\beta QE - (1-\beta) Data > maxUtility$ \textbf{then}} 
               \State {\indent \hspace{9mm} $maxUtility=\beta QE - (1-\beta) Data$;}
              \State {\indent \hspace{9mm} $r_{ik}^{(t)}=r$;}
            \EndFor
         \EndIf
     \State{\textbf{Update} weighting parameters $\rho$, $\omega$ $\gamma$;}
     \State \parbox[t]{\dimexpr\linewidth-\algorithmicindent}{\textbf{Compute} $AQ_{i}$, $E_{i}$, $F_{i}$ and $U_{i}$ according to respectively (\ref{eq1}), (\ref{eq5}), (\ref{eq6}) and (\ref{obj});}
       \State{\textbf{Update} $B_{i}^{(t)}$;}    
       \If{$B_{i}^{(t)} = B_{i}^{max}$}
            \State \parbox[t]{\dimexpr\linewidth-\algorithmicindent}{$BufferStatus=True$; \hspace{1mm} $L_{i}=t-A_{i}$;}
      \EndIf
       \If{$(v_{i},\lceil (t-A_{i})/C \rceil,r_{ik}^{(t)}) \notin M_{k}^{(t)}$}
         \State \parbox[t]{\dimexpr\linewidth-\algorithmicindent} 
           {$d_{ik}^{(t)}=1$;  \hspace{1mm}  $BT_{i}=BT_{i}+r_{ik}^{(t)}$; \strut}
       \EndIf
\State {\textbf{Return} $U_{i}$;}
\end{algorithmic}
\begin{tabular}{p{7.6 cm}}
\hline \\
\end{tabular}

After adjusting the weights, the client buffer level and the overall incurred backhaul network data traffic are accordingly updated. It is noted that in the startup phase, the buffer level is updated knowing that the content is not being consumed by the media player in this phase. If there are new requested chunks downloaded from the backhaul server, the algorithm then proceeds with updating the cache contents using our proposed heuristic which is described in the next section.  

\subsection{Retention-based Cache Replacement Heuristic (RBCRH)} 
\label{sec:cacheupdating}

The most beneficial cache update method will minimize the number of cache misses when clients request video chunks. As pointed out in \cite{Liu2016}, the proactive cache replacement becomes more challenging when the clients mobility are taken into account due to variation in the channel quality and the requested bitrates during different time periods.  \\ \\ 
\begin{tabular}{p{7.6 cm}}
\hline 
\textbf{Subroutine 2:} Steady State \\
\hline \\
\end{tabular}
\begin{algorithmic}[1]
\small    
\vspace{-1em}
        \State {\textbf{Run} the same code lines (4)-(37) as in Startup Phase;} 
\State{\textbf{Update} weighting parameters $\rho$, $\omega$ $\gamma$;}
     \State \parbox[t]{\dimexpr\linewidth-\algorithmicindent}{\textbf{Compute} $AQ_{i}$, $E_{i}$, $F_{i}$ and $U_{i}$ according to respectively (\ref{eq1}), (\ref{eq5}), (\ref{eq6}) and (\ref{obj});}
     \State{\textbf{Update} $B_{i}^{(t)}$;}    
       \If{$(v_{i},\lceil (t-A_{i})/C \rceil,r_{ik}^{(t)}) \notin M_{k}^{(t)}$}
         \State \parbox[t]{\dimexpr\linewidth-\algorithmicindent} 
           {$d_{ik}^{(t)}=1$;  \hspace{1mm}  $BT_{i}=BT_{i}+r_{ik}^{(t)}$; \strut}
       \EndIf
\State {\textbf{Return} $U_{i}$;}
\end{algorithmic}
\begin{tabular}{p{7.6 cm}}
\hline \\
\end{tabular}

Since the future arrival/departure or the requested chunks by the clients are not known in advance, we design a probabilistic heuristic that relies on two separate information: 1) 
Video viewing statistics, namely per-video retention, which we assume to be continuously collected and available from origin service provider (e.g., YouTube collects such stats as "audience retention" but it is currently only available for video owner), 2) current clients' bitrate allocation history. 
The rationale behind using retention is to have prior knowledge on viewing behavior to help with limited size caches and cold cache situations. While the statistics help prioritize video segments, we use the clients' bitrate history to prioritize between different representations of specific segments. Contrary to viewer retention, the latter information cannot be learned in advance but is determined by current and recent system state, hence the cache maintains this information. Our cache updating strategy computes a probability for clients to request a specific video chunk using these two sources of information each time the cache needs to make a decision on which chunk, if any, to evict.   

Once the cache update procedure is called at each edge server, the heuristic computes \textit{weight} and \textit{value} for each set of chunk/bitrate of different requested videos at the current time slot. The weight (occupied space in the cache) of the video chunk is equal to the multiplication of chunk size and its allocated bitrate. The value of a chunk is a unitless quantity describing the probability that it will be requested in the future. This value is computed considering all those clients streaming from the same edge server that are currently downloading earlier chunks of the same video. In other words, these clients will request the chunk in question, or another one with same index but different bitrate, unless they abandon viewing the video before that point of time. The calculation needs to consider both the index of the chunk as well as its bitrate: For each of those clients, the heuristic first computes the likelihood that the client will be still active in its streaming session when it reaches the point of video corresponding to the chunk whose value is being computed. Second, it measures how frequently the bitrate of that chunk has been accessed by each of these other clients during the past. 

Mathematically speaking, assume that the client $i$ has been allocated to server $k$ at time slot $t$ i.e. $a_{ik}^{(t)}=1$. For computing the value of its current chunk with index $\lceil \frac{t-A_{i}}{C} \rceil$, the heuristic first creates the following list of other potential clients which are the clients allocated to the same server as client $i$ and streaming earlier chunks of the same video: 
\begin{align}
\label{eq8}
S=\{j \hspace{1mm} | \hspace{1mm} a_{jk}^{(t)}=1, v_{j}=v_{i}, \lceil \frac{t-A_{j}}{C} \rceil \leq \lceil \frac{t-A_{i}}{C} \rceil\}
\end{align}
The likelihood of caching the chunk of client $i$ allocated to server $k$ at time slot $t$, $P_{cache}(i,k,t)$, is then computed using the following union probability:    
\begin{align}
\label{eq9}
P_{cache}(i,k,t)=P((\cup A_{(j,i,k,t)} \hspace{2mm} \forall j \in S) \cup A_{(new,i,k,t)})  
\end{align}
where for each client $j \in S$ at time slot $t$, the notation $A_{(j,i,k,t)}$ denotes the event that client $j$ has perceived the bitrate of client $i$'s chunk during the past time slots and will be still active in its session until it reaches the client $i$'s chunk. In order to consider the effect of clients' arrival/departure interleaving, we also take into account the arrival of a new client in the caching probability computation. The similar event $A_{(new,i,k,t)}$ is also considered for the new arrival. The probability of the corresponding event for each client $j \in S$ is derived using the following multiplication: 
\begin{align}
\label{eq10}
P(A_{(j,i,k,t)})=P_{reach}(j,i,k,t) \times P_{acc}(j,i,k,t)
\end{align}
where the first term, $P_{reach}(j,i,k,t)$, is the probability that the client $j \in S$ will not abandon the stream before reaching the chunk of th client $i$. We note that in the computation of the caching probability, we assume that all clients in set $S$ stream the video continuously without interruption, however, our methodology can be easily adapted to the case when some of the clients may jump form one part of the video to the later parts. This adaptation can be done by dynamically updating the set of clients in $S$ at each time slot and computing the caching value, accordingly.
Assuming that the retention function $P_{act}$, which specifies the probability of a newly arrived client to view different parts of the video, is known and provided by the origin video server, the probability $P_{reach}(j,i,k,t)$ is therefore estimated using the following relation:
\begin{align}
\label{eq11}
P_{reach}(j,i,k,t) \approx 1-(P_{act}(\lceil \frac{t-A_{j}}{C} \rceil)-P_{act}(\lceil \frac{t-A_{i}}{C} \rceil))
\end{align}
If a given client is currently downloading a chunk of video before the chunk index of client $i$, 
it is evident that the closer it is to the chunk index of client $i$  
the higher probability that client $j$ will request that other chunk in the future. For the second criteria, the heuristic measures the probability $P_{acc}(j,i,k,t)$ which states how frequently the bitrate of the considered chunk of client $i$ has been accessed by client $j$ during the past time slots. This quantity is obtained using the following relation: 
\begin{align}
\label{eq12}
P_{acc}(j,i,k,t)=\frac{\sum_{t^{\prime}=A_{j}}^{t} a_{jk}^{(t^{\prime})} \cdot \mathbb{I}(r_{jk}^{(t^{\prime})}=r_{ik}^{(t)})}{\sum_{t^{\prime}=A_{j}}^{t} a_{jk}^{(t^{\prime})}}
\end{align} 
where the identity function $\mathbb{I}(r_{jk}^{(t^{\prime})}=r_{ik}^{(t)})=1$ if the allocated bitrate to client $j$ on server $k$ at time slot $t^{\prime}$ is equal to the bitrate of client $i$ at time slot $t$ and $\mathbb{I}(r_{jk}^{(t^{\prime})}=r_{ik}^{(t)})=0$, otherwise.   \\ \\ 
\begin{tabular}{p{7.6 cm}}
\hline 
\textbf{Subroutine 3:} Retention-based Cache Replacement Heuristic (RBCRH) (Run by the edge server)   \\
\hline \\
\end{tabular}
\begin{algorithmic}[1]
\small    
\vspace{-1em}
\For{each edge server $1 \leq k \leq K$}
  \If{$d_{ik}^{(t)}==1$, \hspace{1mm} for at least one client $i: a_{ik}^{(t)}=1$} 
\State \parbox[t]{\dimexpr\linewidth-\algorithmicindent}{\textbf{Set} $V,I,W,P=$ Lists of respectively the \\
videos, chunk indexes, the weights and the \\ 
updated values of the existing chunks in cache;  \strut} 
       \For{each each client $i$ such that $a_{ik}^{(t)}==1$}
           \State  \parbox[t]{\dimexpr\linewidth-\algorithmicindent}{\textbf{if} $v_{i} \notin V$ \textbf{OR} $\lceil (t-A_{i})/C \rceil \notin I$  \\
           \indent \hspace{3mm} \textbf{OR} $r_{ik}^{(t)} \cdot C \notin W$ \textbf{then} \strut}
           \State  \parbox[t]{\dimexpr\linewidth-\algorithmicindent}{\indent \hspace{6mm} \textbf{Create} set $S$ from (\ref{eq8}); \strut}   
           \State  \parbox[t]{\dimexpr\linewidth-\algorithmicindent}{\indent \hspace{6mm} \textbf{Compute} $P_{reach}(j,i,k,t)$ and $P_{acc}(j,i,k,t)$ \\
           \indent \hspace{6mm} from (\ref{eq11}) and (\ref{eq12}) for each client $j \in S$; \strut} 
\State  \parbox[t]{\dimexpr\linewidth-\algorithmicindent}{\indent \hspace{6mm} \textbf{Compute} $P_{cache}(i,k,t)$ value for client $i$ \\
\indent \hspace{6mm} according to relation (\ref{eq9}); \strut} 
\State  \parbox[t]{\dimexpr\linewidth-\algorithmicindent}{\indent \hspace{6mm} \textbf{Append} video $v_{i}$, chunk index \\ 
\indent \hspace{6mm} $\lceil (t-A_{i})/C \rceil$ and its weight $C \cdot r_{ik}^{(t)}$  \\   
\indent \hspace{6mm} to lists respectively $V$, $I$ and $W$; \strut} 
\State \parbox[t]{\dimexpr\linewidth-\algorithmicindent}{\indent \hspace{6mm} \textbf{Append} $P_{cache}(i,k,t)$ value to $P$}
       \EndFor
       \State \parbox[t]{\dimexpr\linewidth-\algorithmicindent}{\textbf{Sort} list $P$ and accordingly lists $V,I,W$ in \\
       \indent \hspace{6mm} decreasing order of caching values; \strut} 
       \State {$sumofWeights=0$;} 
       \For{$index=1$ to $size(P)$}
            \State \parbox[t]{\dimexpr\linewidth-\algorithmicindent}{$sumofWeights= \\
            \indent \hspace{18mm} sumofWeights+W(index)$; \strut}   
           \If{$sumofWeights > Q$}
                \State{\textbf{Break};}
           \EndIf
           \State \parbox[t]{\dimexpr\linewidth-\algorithmicindent}{$M_{k}^{(t)}=M_{k}^{(t)} \cup \\
          \indent \hspace{16mm} (V(index),I(index),R(index))$;} 
       \EndFor
     \EndIf
\EndFor
\end{algorithmic}
\begin{tabular}{p{7.6 cm}}
\hline \\
\end{tabular}

Similarly, the likelihood that the new arrival reaches the current chunk of client $i$ is given by:
\begin{align}
\label{eq13}
P_{reach}(new,i,k,t)=1-(1-P_{act}(\lceil (t-A_{i})/C \rceil))
\end{align}
Since a new arrival has not yet started the video streaming, we estimate the probability that the new arrival has accessed the same bitrate as client $i$ during the past time slots with the equal probability $P_{acc}(new,i,k,t) \approx 1/|R|$, where $R$ is the set of available bitrates on the origin server.

It should be noted that we assume the events $A_{(j,i,k,t)}$ and $A_{(new,i,k,t)}$ in relation (\ref{eq9}) to be independent ignoring the complex interdependencies among the clients for the sake of analytical simplicity.  
At each edge server, the heuristic then sorts the current chunks in decreasing order of their computed caching values $P_{cache}$. In the sorted order, the chunks are then inserted into the cache until the sum weights of the chunks exceed the cache capacity. The pseudo-code of the proposed cache updating heuristic named as retention-based cache replacement heuristic (RBCRH) has been shown in Subroutine 3.

\subsection{Computational Complexity}
\label{subsec:computationalcomplexity}

In this section, we derive the worst case computational complexity of bitrate allocation algorithm CSBAA and the cache replacement heuristic RBCRH, separately. It is seen from Algorithm 1 that within $|T|$ time slots, for each mobile client, the allocation of the client to the most appropriate edge server takes the worst case complexity of order $O(K)$. If the client is downloading in the middle of its current chunk, the same bitrate as the previous time slot is allocated to the client with complexity of $O(1)$. If the client is at the beginning of the new chunk, depending on its buffer status, either the startup phase or the steady state is run with the same order of complexity. 

According to the Startup phase, the computation of $estThr$ from the previous chunk performs with the worst case complexity of $C \cdot S$, where $C$ and $S$ are the fixed chunk size and the number of clients, respectively. Also, estimating the switching threshold $\delta_S$ based on the buffer level of the client takes $O(|R|)$ time. 
With $|R|$ available bitrates, finding the most sustainable bitrate which satisfies the thresholds and yields the maximum utility value for the objective function ($\ref{obj}$) performs with the complexity of order $O(|R|^2)$. It is noted that updating each weighting parameter of QoE term in the objective function requires $O(|R|)$ time. Putting the above complexities together results in the following worst case time complexity for algorithm CSBAA:  
\begin{align}
\label{eq14}
T_{CSBAA} \in O(|T| \cdot S \cdot K \cdot (|R|^2 + C \cdot S))
\end{align}

For the cache updating heuristic, the RBCRH run at each edge server updates the cache contents for $|T|$ number of time slots in the worst case. Since the edge servers run the heuristic independently, we analyze in the following the worst case time complexity of running the heuristic at one server. 

At each edge server, the heuristic first updates the caching value of the current chunks in the cache. With the fixed cache size $Q$ at each edge server, the number of available chunks from different videos and with multiple qualities in the cache can be maximum $Q/Cr_{min}$, where $r_{min}$ is the minimum available bitrate in set $R$. For each of these chunks, the computation of caching value requires finding the set of clients $S$ according to ($\ref{eq8}$) which takes $O(S)$, the computation of $P_{reach}$ with the complexity of $O(1)$, the estimation of accessibility probability $P_{acc}$ from ($\ref{eq12}$) with the complexity of $O(|T|)$ and finally, deriving the caching value $P_{cache}$ according to ($\ref{eq9}$) which requires $O(S)$ time. Therefore, the first step takes $O((Q/Cr_{min}) \cdot S \cdot (|T|+1))$ time to perform. 

The heuristic then computes the caching value for the new requested chunks by the clients. In the worst case, all $S$ clients are allocated to one server and also, none of the requested chunks by $S$ clients exist in the cache. Therefore, the overall time complexity of value computation for the downloaded chunks will be $O(S(S(|T|+1)))$. Together with the current chunks, the overall time of caching value computation will be in the order of $O(S \cdot |T|)(Q/Cr_{min}+S)$.

The heuristic then sorts all $Q/Cr_{min}+S$ chunks based on their caching values in the decreasing order with the computation time of $O(Q/Cr_{min})log(Q/Cr_{min})$ and puts them in the sorted order into the cache within $Q/Cr_{min}+S$ time. Combining the computation times of the above three steps and for $|T|$ number of time slots, the following worst case time complexity is obtained for running the RBCRH at each edge server:   
\begin{align}
\label{eq15}
T_{RBCRH} \in O(&|T| \cdot ((Q/Cr_{min}+S) \nonumber  \\
& \cdot (S \cdot |T| + log(Q/Cr_{min}+S))))
\end{align}

It is also noteworthy to mention that one of the important factors that can assist the designers of edge caching adaptive video streaming architecture is the optimal sizing of the cache at the edges. However, under multiple video streaming scenario and varying wireless link quality of the mobile clients, deriving the optimal cache size would be complicated which demands for future investigation.

\section{Simulation Results}
\label{sec:results}

In this section, we evaluate the performance of mobile edge caching adaptive video streaming with the radio access link level traces generated using SimuLTE \cite{SimuLTE}. Our objectives are to quantify the benefits of joint QoE-traffic optimization in two different scenarios: 1) When the set of video chunks cached at edge servers is fixed during the whole video streaming process. 2) RBCRH cache updating heuristic is invoked which updates the edge cache every time a new chunk of video is downloaded from the origin server. 


\begin{table}[t]
\label{table:simulte}
\caption{Simulation parameters and their values.}
\centering
\begin{tabular}{|p{4cm}|p{3cm}|}
\hline  
\textbf{Simulation Parameter}  &  \textbf{Corresponding Value}   \\
\hline\hline
\hspace{2mm} Number of clients &  \multicolumn{1}{|l|}{\emph{100}}    \\
\hspace{2mm}  Number of eNodeBs  &   \multicolumn{1}{|l|}{\emph{10}}     \\
\hspace{2mm}  Number of time slots   &   \multicolumn{1}{|l|}{300}  \\
\hspace{2mm}  Time slot duration   &   \multicolumn{1}{|l|}{\emph{1 seconds}}  \\
\hspace{2mm}  Chunk size   &   \multicolumn{1}{|l|}{\emph{5 seconds}}  \\
\hspace{2mm}  Client Buffer Capacity   &   \multicolumn{1}{|l|}{\emph{250 Mb}}  \\
\hspace{2mm}  Cache size   &   \multicolumn{1}{|l|}{\emph{2 Gb}}  \\
\hspace{2mm}  Fairness Threshold ($\delta_{F}$)   &   \multicolumn{1}{|l|}{0.5}  \\
\hline 
\hspace{2mm} Client antenna gain    &   \multicolumn{1}{|l|}{\emph{0 dBi}}   \\
\hspace{2mm} eNodeB antenna gain    &   \multicolumn{1}{|l|}{\emph{18 dBi}}  \\  
\hspace{2mm} Client speed   &    \multicolumn{1}{|l|}{\emph{8.33 mps}}  \\
\hspace{2mm}  Maximum Transmission   & \multicolumn{1}{|l|}{\emph{26 dBm}}   \\
\hspace{4mm}  power per client   & \multicolumn{1}{|l|}{}    \\
\hspace{2mm}  Channel bandwidth  & \multicolumn{1}{|l|}{\emph{5 MHz}}  \\ 
\hspace{2mm}  Number of downlink RBs  & \multicolumn{1}{|l|}{\emph{28}}  \\  
\hspace{2mm}  Scheduler  &    \multicolumn{1}{|l|}{\emph{Proportional Fairness}}    \\
\hspace{2mm}  Channel model   &   \multicolumn{1}{|l|}{\emph{Urban Macrocell}}  \\
\hspace{2mm}  Shadowing    &   \multicolumn{1}{|l|}{\emph{Disabled}}  \\
\hline
\end{tabular}
\label{table:simulationparameters}
\end{table} 

\subsection{Simulation Setup}
\label{subsec:setup}

We simulate DASH clients during 300 time slots i.e. $|T|=300$ with duration of $\Delta t=1$ second for each time slot. CSBAA algorithm is implemented in a Matlab simulator and its performance is evaluated with the radio access link level traces from LTE simulator \cite{SimuLTE}. A fairness threshold $\delta_{F}=0.5$ is considered in all simulations. The network setup includes 100 UEs (mobile clients) which are associated to 10 eNodeBs according to their physical proximity. 
We consider our simulations based on a scenario in which the mobile clients commute to their work by riding a bus or vehicle with constant speed of 8.33 \emph{m/s} \cite{Essaili2015} and with the linear mobility pattern. Under such mobility scenario, the downlink SNR values of the clients from every eNodeB during 300 time slots are obtained following the LTE wireless link specifications reported in \cite{LTEThroughput}.


The chunks of four videos with different popularities are available in ten different qualities 
$[15,17,22,26,30,35,38,43,45,50 Mbps]$ at the backhaul server. Each video has a duration of 270$s$ and each chunk has length $C=5s$. The buffer of each mobile client has the constant capacity of 250 $Mb$ video data and, unless explicitly mentioned, the time that each client start its streaming session (arrival time slot) is drawn from the uniform interval $U[0,30s]$. A linear retention curve is considered for all of the clients unless otherwise mentioned.
More precisely, in order to prioritize the videos based on their popularity, we assume that after the client starts its streaming session, it remains active until the first $90s$, $50s$, $50s$ and $30s$ of the video, if the requested video is respectively the first, second, third or the fourth one. The client then departs from its session at a time slot which is chosen from the rest of the time slots until the end of the video.  


The cache size of each edge server is 2 $Gb$ 
and with total available per slot bandwidth $5 MHz$, the number of 28 LTE downlink resource blocks are also available per time slot at each base station. We should also note that for each part of simulation, the average of the results taken over 20 runs of simulation with confidence interval of 95\% are presented. We have summarized the list of simulation parameters and their corresponding values in Table. \ref{table:simulationparameters}.

\begin{figure}[t] 
  \begin{subfigure}[b]{0.49\linewidth}
    \centering
     \includegraphics[width=1\linewidth]{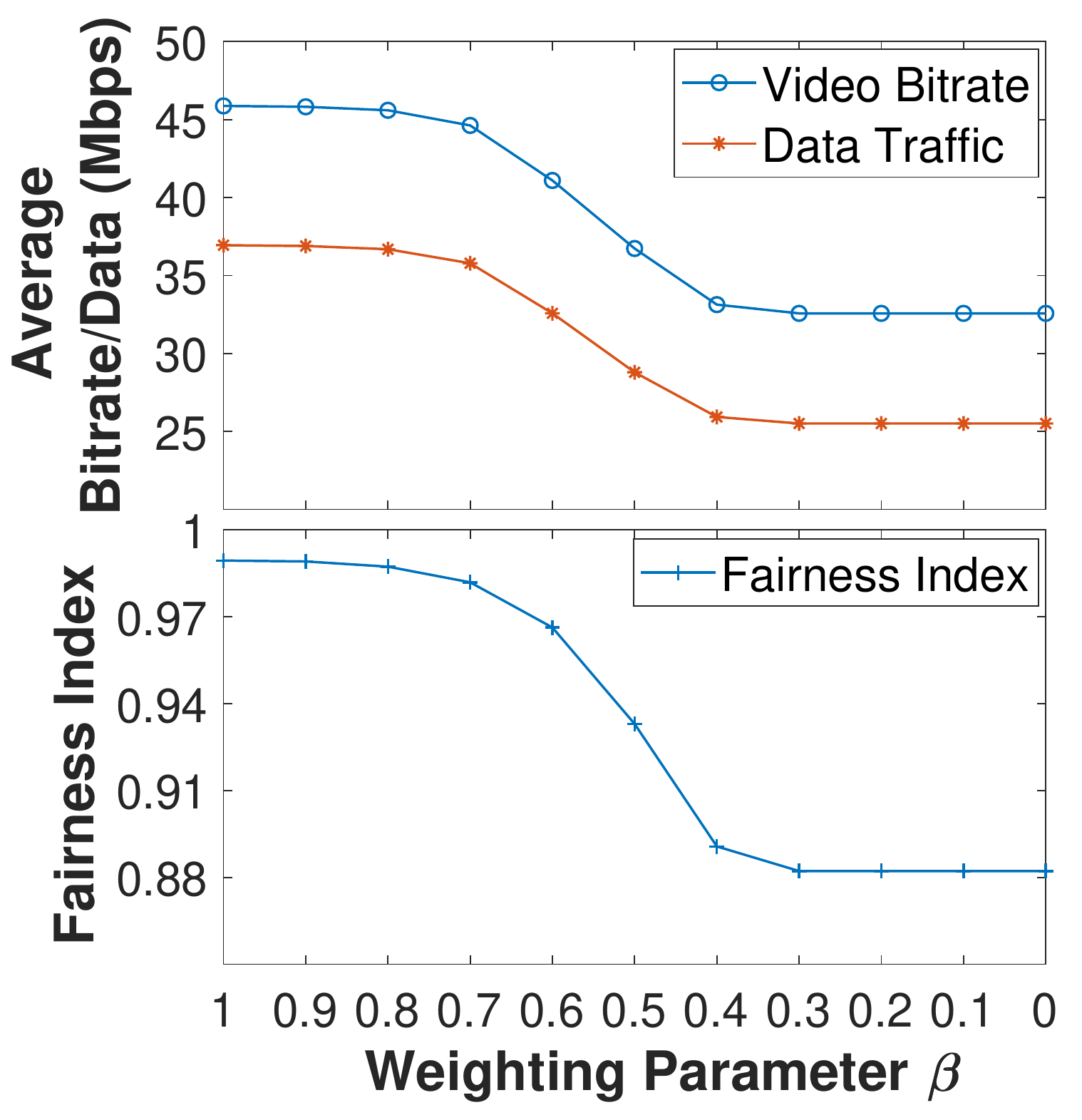}  
    \caption{Video Data and Fairness} 
    \label{fig:averagebitratefairness}
    \vspace{0.2ex}
  \end{subfigure}
  \begin{subfigure}[b]{0.49\linewidth}
    \centering
   \includegraphics[width=1\linewidth]{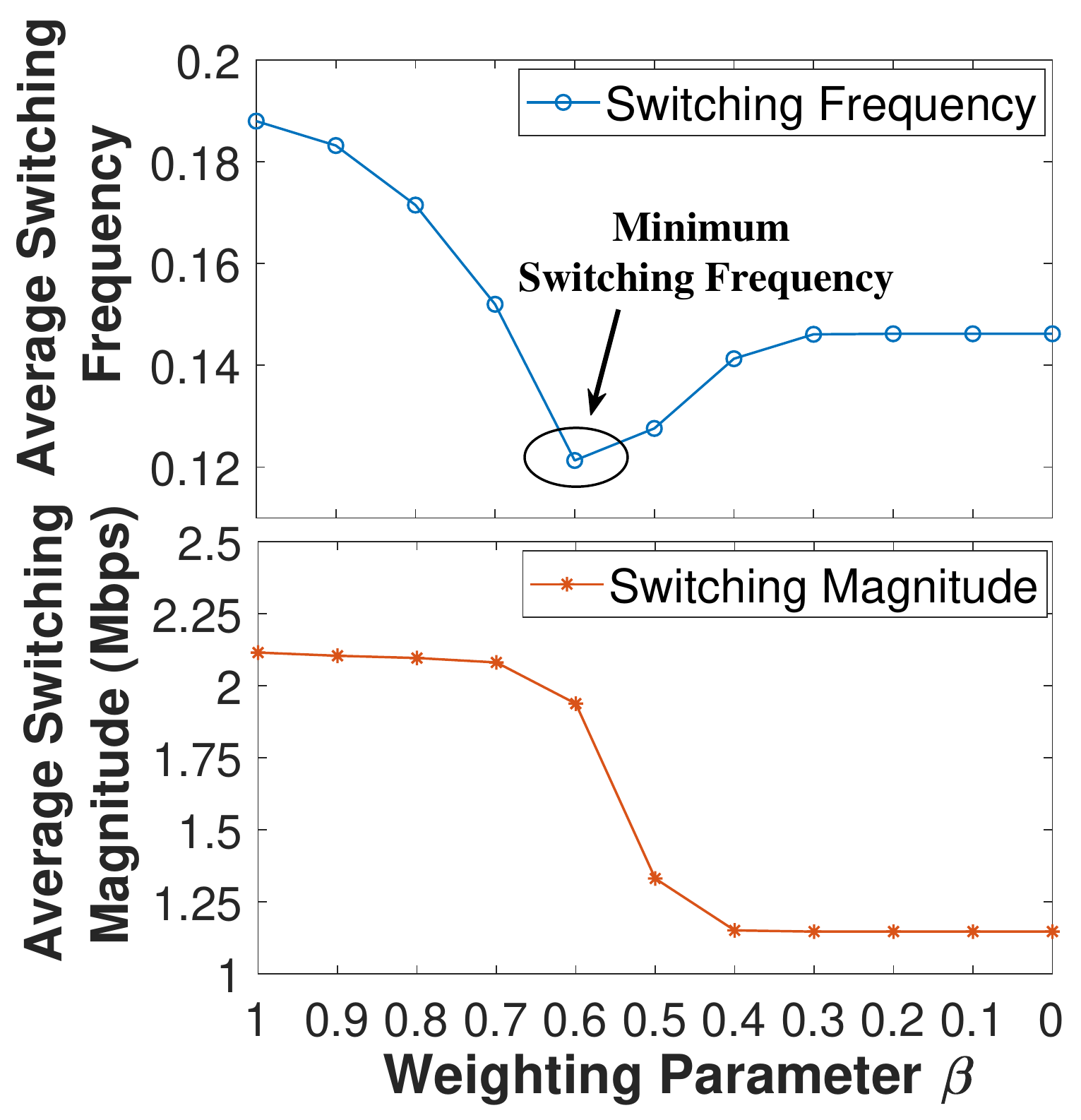} 
    \caption{Switching} 
    \label{fig:switchingfrequencymagnitude}
    \vspace{0.32ex}
  \end{subfigure} 
  \caption{Impact of QoE-traffic weighting parameter $\beta$ on a) average bitrate/data and fairness index and b) switching frequency/magnitude.}  
\end{figure}

\subsection{Tuning the QoE-traffic Trade-off}
\label{subsec:qoetraffictuning}

We first look at the impact of the weight parameter $\beta$ that controls the QoE-traffic trade-off. 
Fig. \ref{fig:averagebitratefairness} shows the impact of decreasing $\beta$ on the average video bitrate/backhaul data traffic and the fairness index. As expected, the average bitrate per client decreases as QoE weighting decreases while it also results in reduction in backhaul data traffic. As the clients will get lower bitrates when most of the chunks are downloaded from the edge servers by decreasing $\beta$, the fairness index among the clients decreases as confirmed from the result in Fig. \ref{fig:averagebitratefairness}.

Increasing the value of the data traffic weight reduces downloading from the origin server. As the results in Fig. \ref{fig:switchingfrequencymagnitude} shows, this in turn reduces bitrate switching as well. However, the impact of decreasing $\beta$ on switching frequency reveals interesting behavior as it reaches an optimal lowest point after which it slightly increases and then converges to a stable point. 

The takeaway here is that the weight parameter $\beta$ acts as a control parameter to tune the traffic-QoE trade-off as intended. This parameter can be set by the MNO to select a desired point of operation which yields less traffic or better QoE. 

\begin{figure}[t] 
  \begin{subfigure}[b]{0.49\linewidth}
    \centering
     \includegraphics[width=1\linewidth]{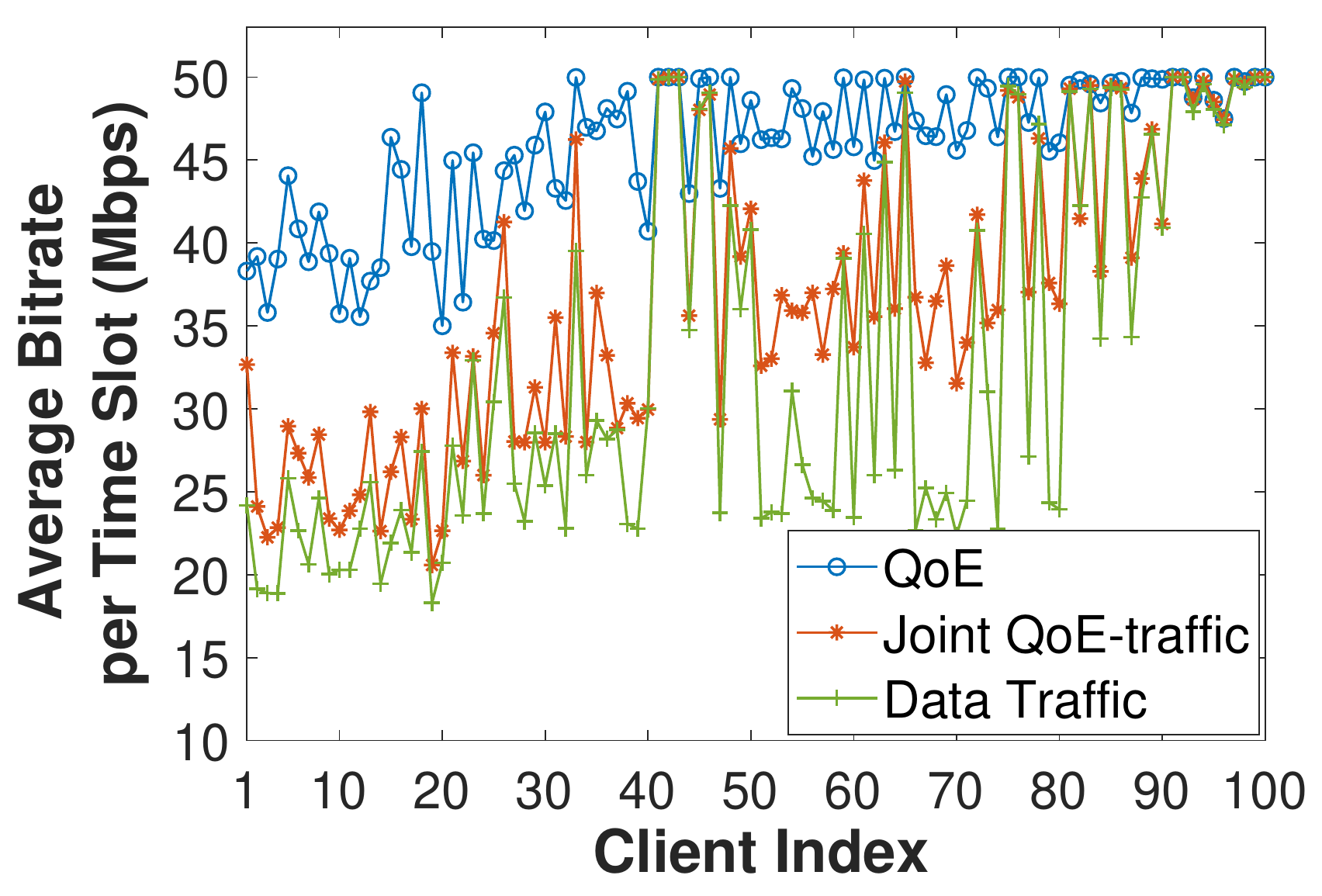} 
    \caption{Average Bitrate} 
    \label{fig:averagebitrate}
    \vspace{0.2ex}
  \end{subfigure}
  \begin{subfigure}[b]{0.49\linewidth}
    \centering
   \includegraphics[width=1\linewidth]{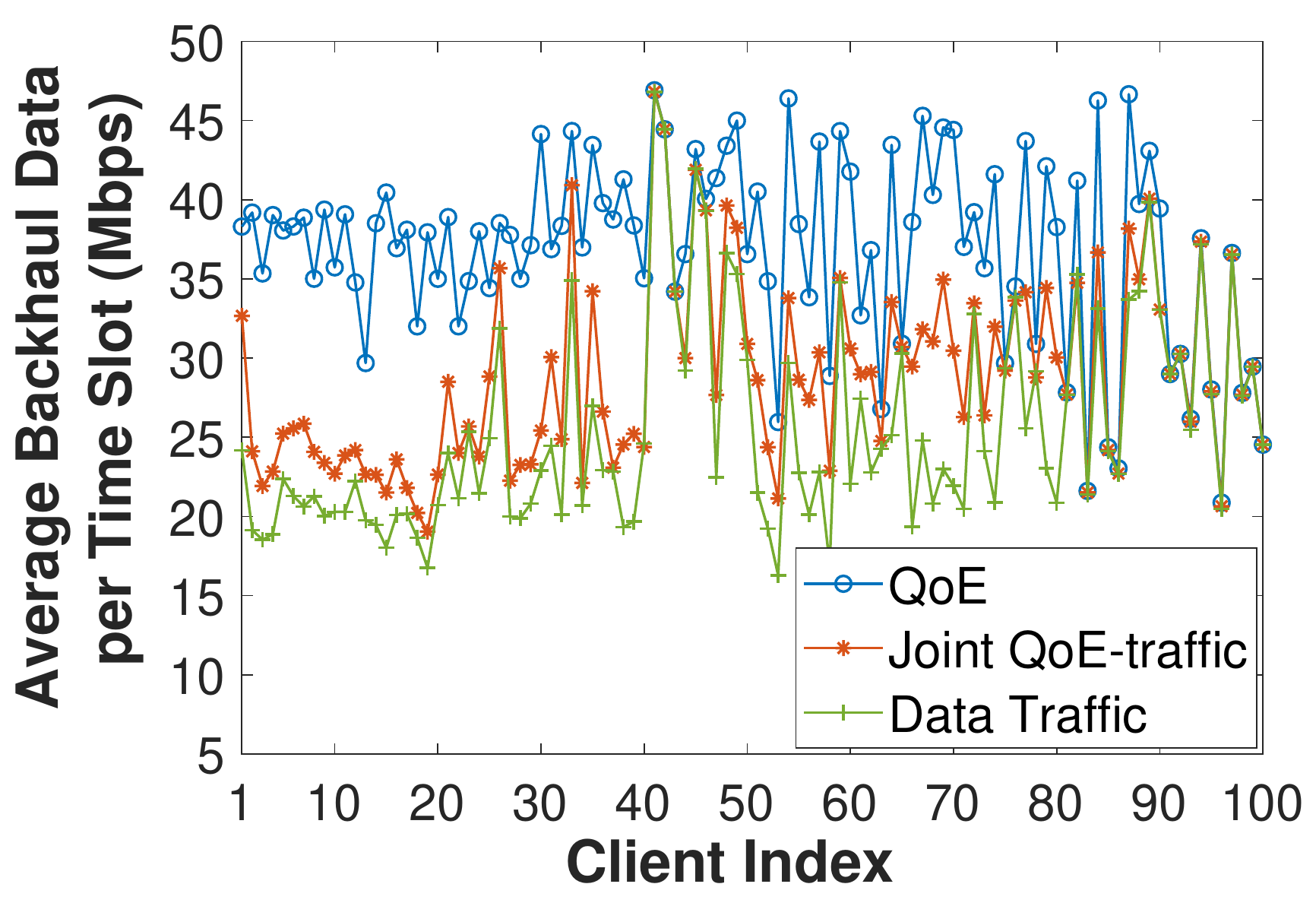} 
    \caption{Average Data Traffic} 
    \label{fig:averagedata}
    \vspace{0.32ex}
  \end{subfigure} 
  \begin{subfigure}[b]{0.49\linewidth}
    \centering
   \includegraphics[width=1\linewidth]{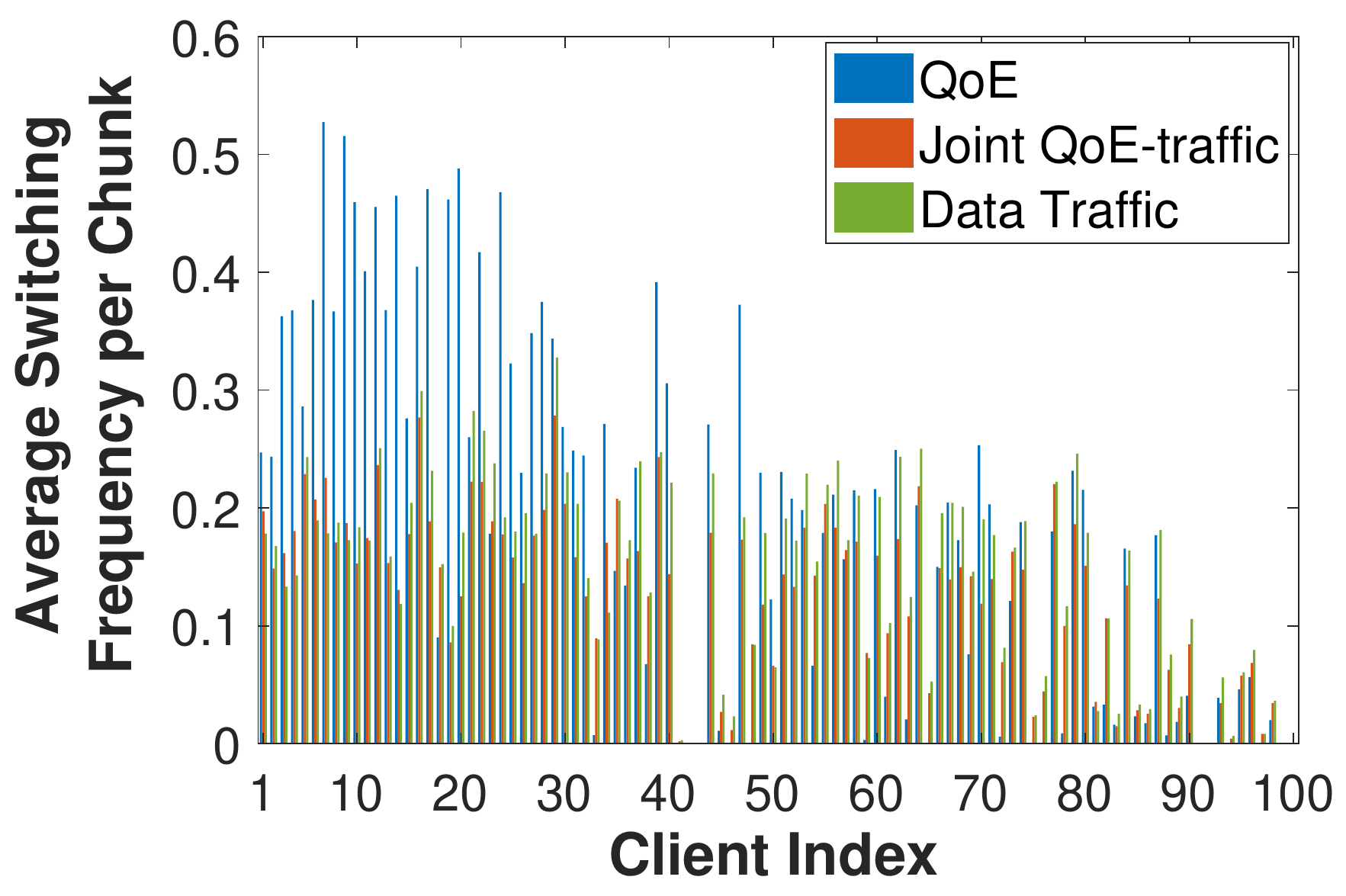} 
    \caption{Switching Frequency} 
    \label{fig:switchingfrequency}
    \vspace{0.12ex}
  \end{subfigure} 
  \begin{subfigure}[b]{0.49\linewidth}
    \centering
   \includegraphics[width=1\linewidth]{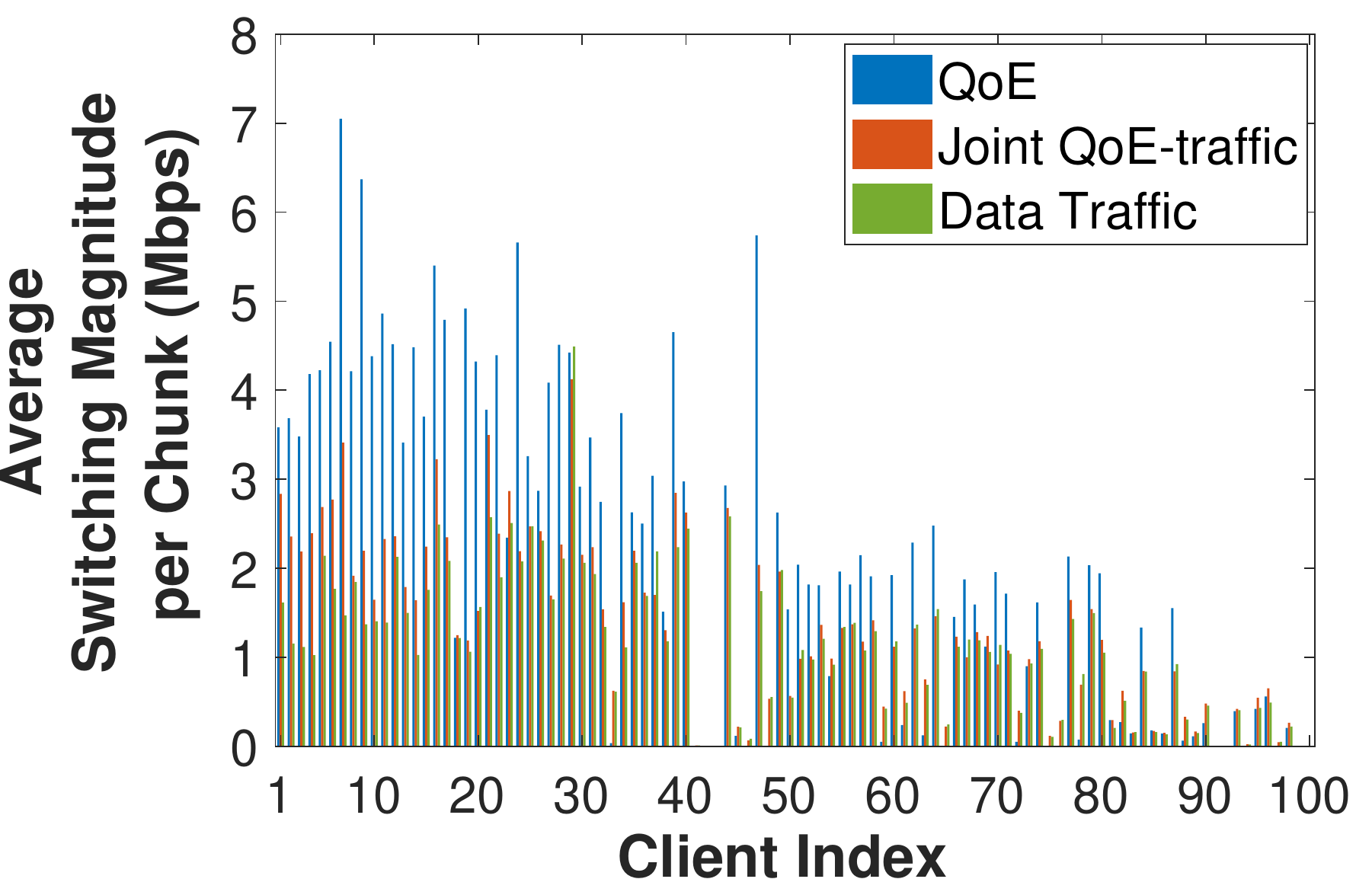}  
    \caption{Switching Magnitude} 
    \label{fig:switchingmagnitude}
    \vspace{0.32ex}
  \end{subfigure} 
  \caption{Comparison between three optimization strategies under fixed cached contents scenario.} 
\end{figure}

\subsection{Joint vs. Single Objective Optimization}
\label{subsec:comparison}

Next we compare the joint optimization to strategies that only try to minimize traffic or maximize QoE:
\begin{itemize}
\item \textbf{QoE Maximization} ($\beta=1$): In this case the mobile clients download chunks with the best possible bitrate, i.e. only maximizing QoE, regardless of the data traffic that it may create on the backhaul network. The chosen chunk will be downloaded from the edge cache if it is there and from the origin server otherwise. 
\item \textbf{Joint QoE-traffic}: In this case, we consider the weight $\beta=0.5$ for the joint optimization problem. Mobile clients download chunks form either the local cache or the origin server such that their obtainable utility (objective value (\ref{obj})) is maximized. 
\item \textbf{Traffic Minimization} ($\beta=0$): In this case, the bitrates are selected so that downloading the corresponding chunks will minimize the backhaul data traffic without considering the resulting QoE. More precisely, if the chunk is available in the edge cache at any bitrate, it is downloaded from there. Otherwise, the chunk is downloaded from the origin server with the lowest bitrate.  
\end{itemize} 

We compare the three strategies under two scenarios: In the first scenario, some uniform randomly chosen chunks are available in the edge caches and they remain unchanged during the whole video streaming session. In the second scenario with the initially empty caches, RBCRH heuristic is employed to update cache contents once new requested chunks are downloaded form the backhaul server. 

\subsubsection{Fixed Cached Contents}


Figure \ref{fig:averagedata} shows that the joint QoE-data traffic optimization reduces the volume of video data downloaded from the origin server compared to the case when the system applies the QoE maximization strategy. However, the average bitrate of the clients noticeably drops with the joint optimization (Figure \ref{fig:averagebitrate}). Further reduction in backhaul data traffic with slight reduction in average bitrate is achieved when the data traffic minimization strategy is taken into account. With the average of the results taken over 20 runs of the simulation,
joint optimization leads to roughly 20\% reduction in both the average bitrate and data traffic compared to QoE maximization. Data traffic minimization strategy reduces the bitrate and traffic by about 30\% on average compared to QoE maximization. 


Although the QoE maximization yields higher average video bitrate for the clients, it leads to higher switching frequency and magnitude compared to the joint optimization or data traffic minimization strategies as observed from the results in Fig. \ref{fig:switchingfrequency} and \ref{fig:switchingmagnitude}.    


\begin{figure}[t] 
  \begin{subfigure}[b]{0.49\linewidth}
    \centering
     \includegraphics[width=1\linewidth]{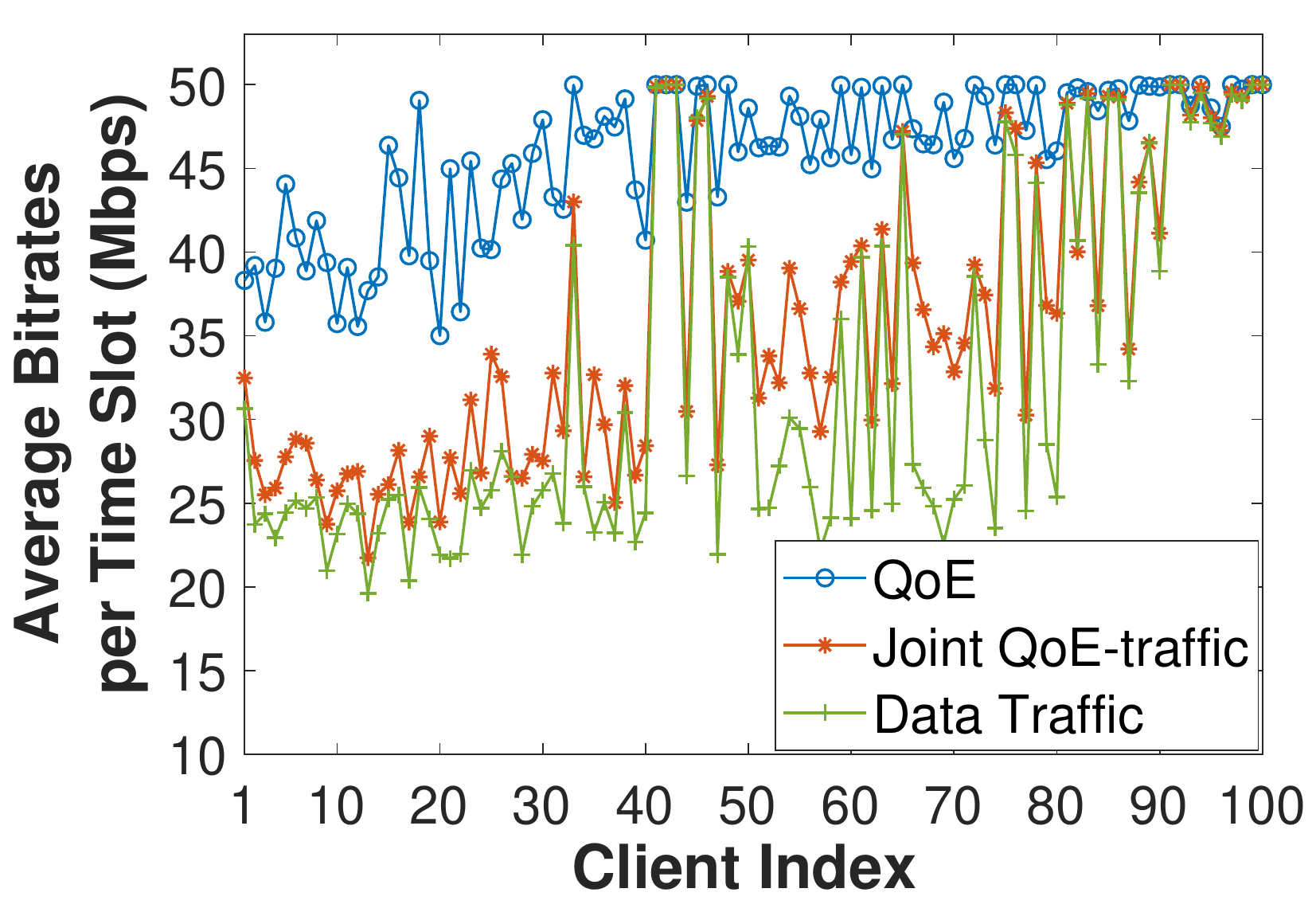}  
    \caption{Average Bitrate} 
    \label{fig:averagebitratecache}
    \vspace{0.2ex}
  \end{subfigure}
  \begin{subfigure}[b]{0.49\linewidth}
    \centering
   \includegraphics[width=1\linewidth]{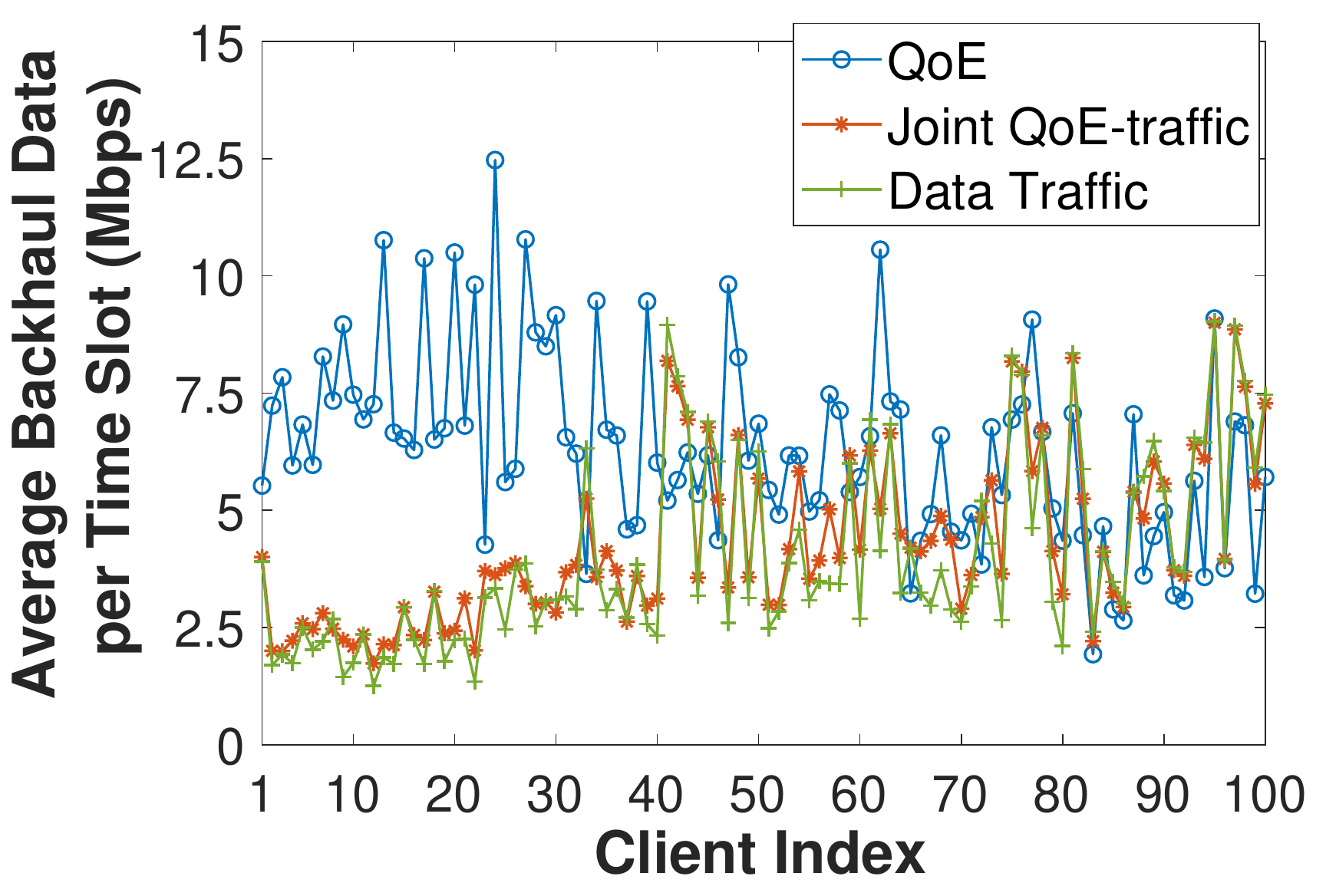} 
    \caption{Average Data Traffic} 
    \label{fig:averagedatacache}
    \vspace{0.32ex}
  \end{subfigure} 
  \begin{subfigure}[b]{0.49\linewidth}
    \centering
   \includegraphics[width=1\linewidth]{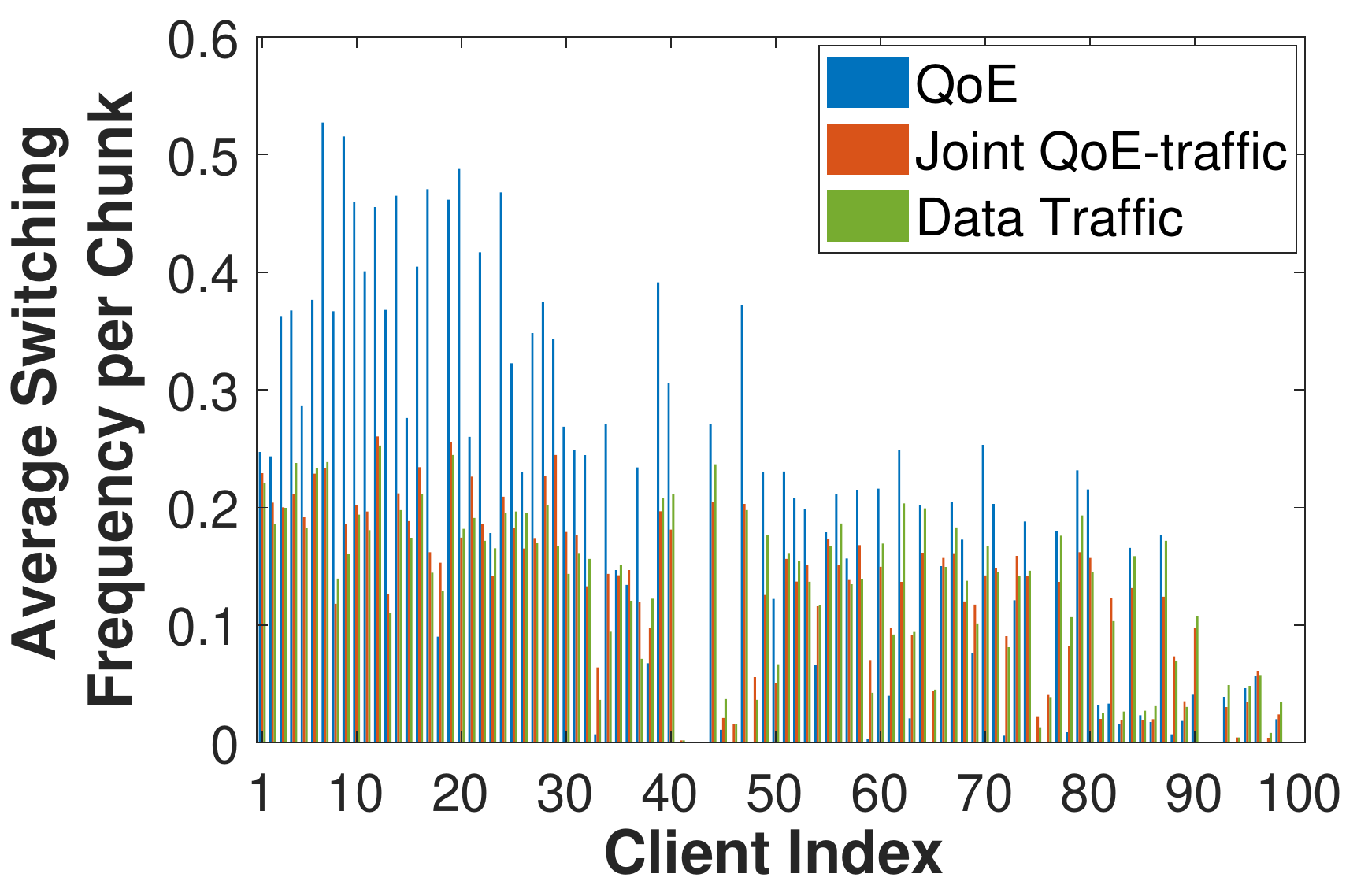} 
    \caption{Switching Frequency} 
    \label{fig:switchingfrequencycache}
    \vspace{0.12ex}
  \end{subfigure} 
  \begin{subfigure}[b]{0.49\linewidth}
    \centering
   \includegraphics[width=1\linewidth]{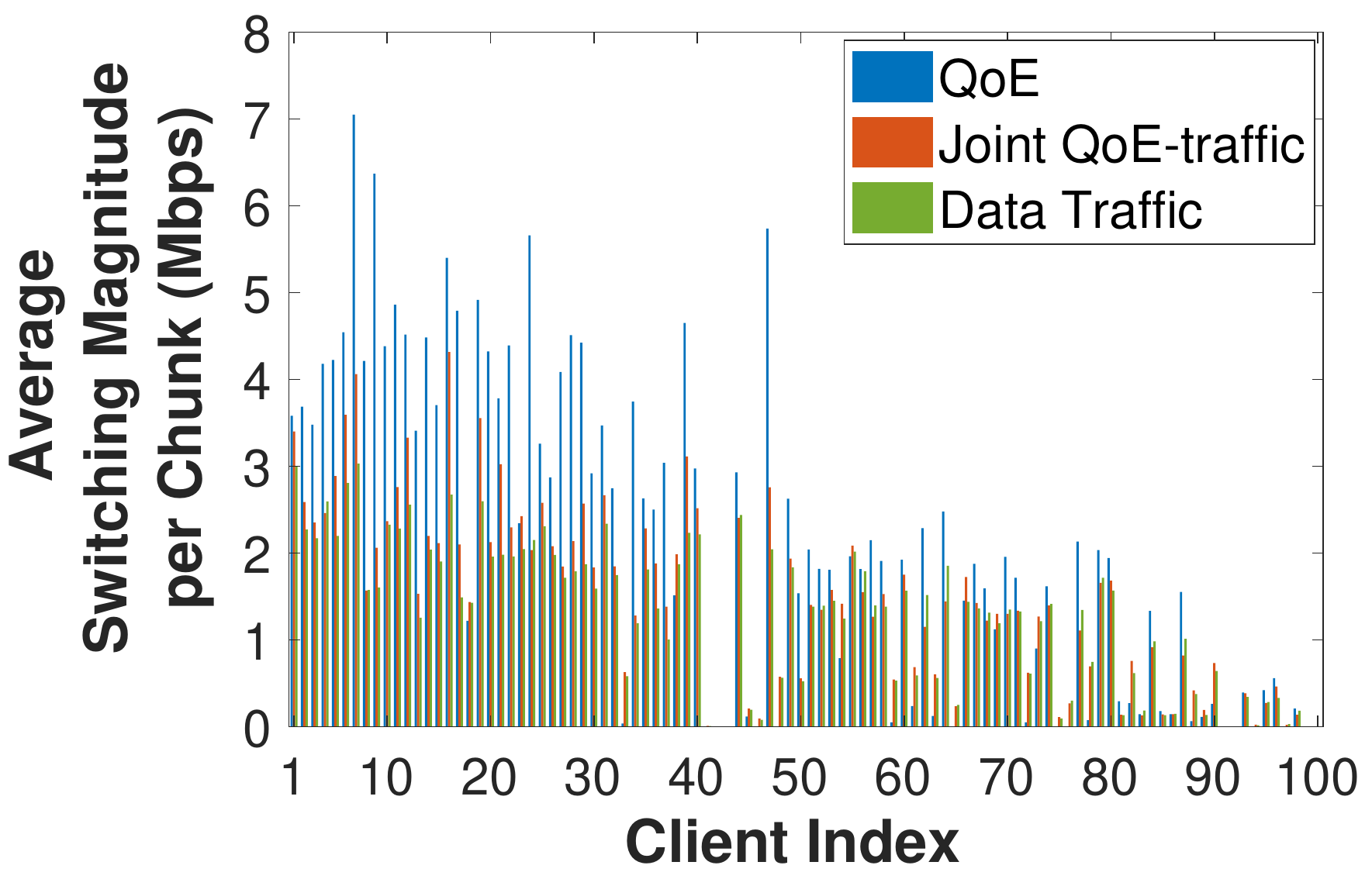}  
    \caption{Switching Magnitude} 
    \label{fig:switchingmagnitudecache}
    \vspace{0.32ex}
  \end{subfigure} 
  \caption{Comparison between three optimization strategies under cache updating scenario.} 
\end{figure}

\subsubsection{Updating Cache Contents}

In the second scenario, with the initially empty cache at the edges, the cache contents are updated using our proposed cache replacement heuristic RBCRH once there are new chunks downloaded from the origin video server and the decision on chunk eviction has to be made. The results of comparing three versions of optimization problem under the second scenario have been shown in \ref{fig:averagebitratecache}-\ref{fig:switchingmagnitudecache}. 

Similar to the first scenario, both joint QoE-traffic optimization and backhaul traffic minimization strategies reduce the volume of downloaded data from the origin server 
compared to QoE maximization. 
With small variation in average video bitrate, updating the cache contents causes higher percentage of traffic reduction compared to the case of fixed cached contents. As the results show, the joint optimization reduces the video bitrate and backhaul traffic by 22\% and 32\%, respectively, compared to QoE maximization, while the data traffic minimization strategy reduces the backhaul traffic by 36\% and video bitrate by 30\% on average. Compared to the first scenario, the backhaul data traffic with the three different strategies drops on average by 80\% with only small differences in the average video bitrates. This observation suggests that updating the cache contents has a more noticeable impact on the backhaul traffic than on the QoE of the clients.

\begin{figure}[t] 
  \begin{subfigure}[b]{0.33\linewidth}
    \centering
     \includegraphics[width=1\linewidth]{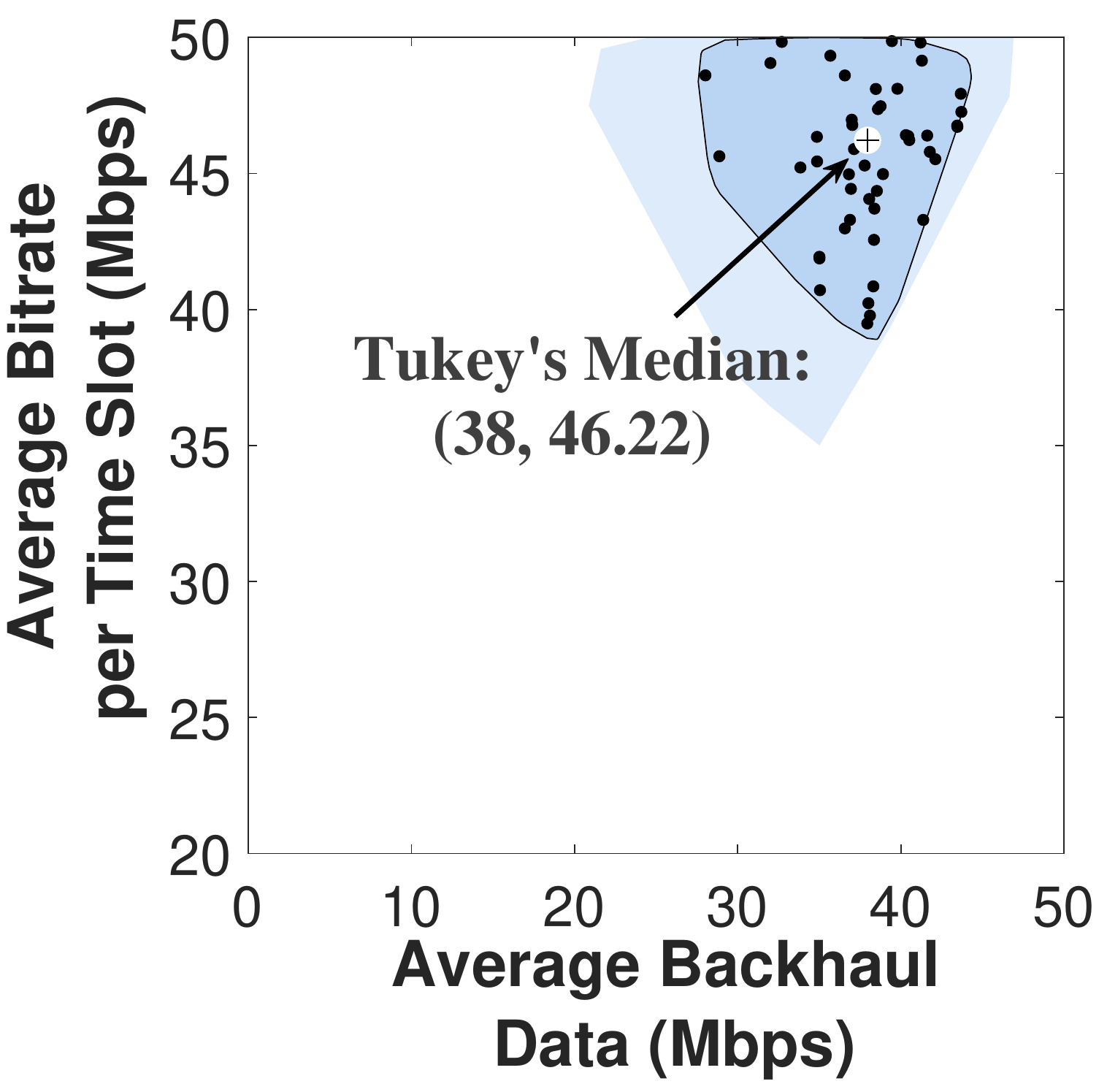} 
    \caption{QoE} 
    \label{fig:bagplotqoefixed}
    \vspace{0.2ex}
  \end{subfigure}
  \begin{subfigure}[b]{0.33\linewidth}
    \centering
   \includegraphics[width=1\linewidth]{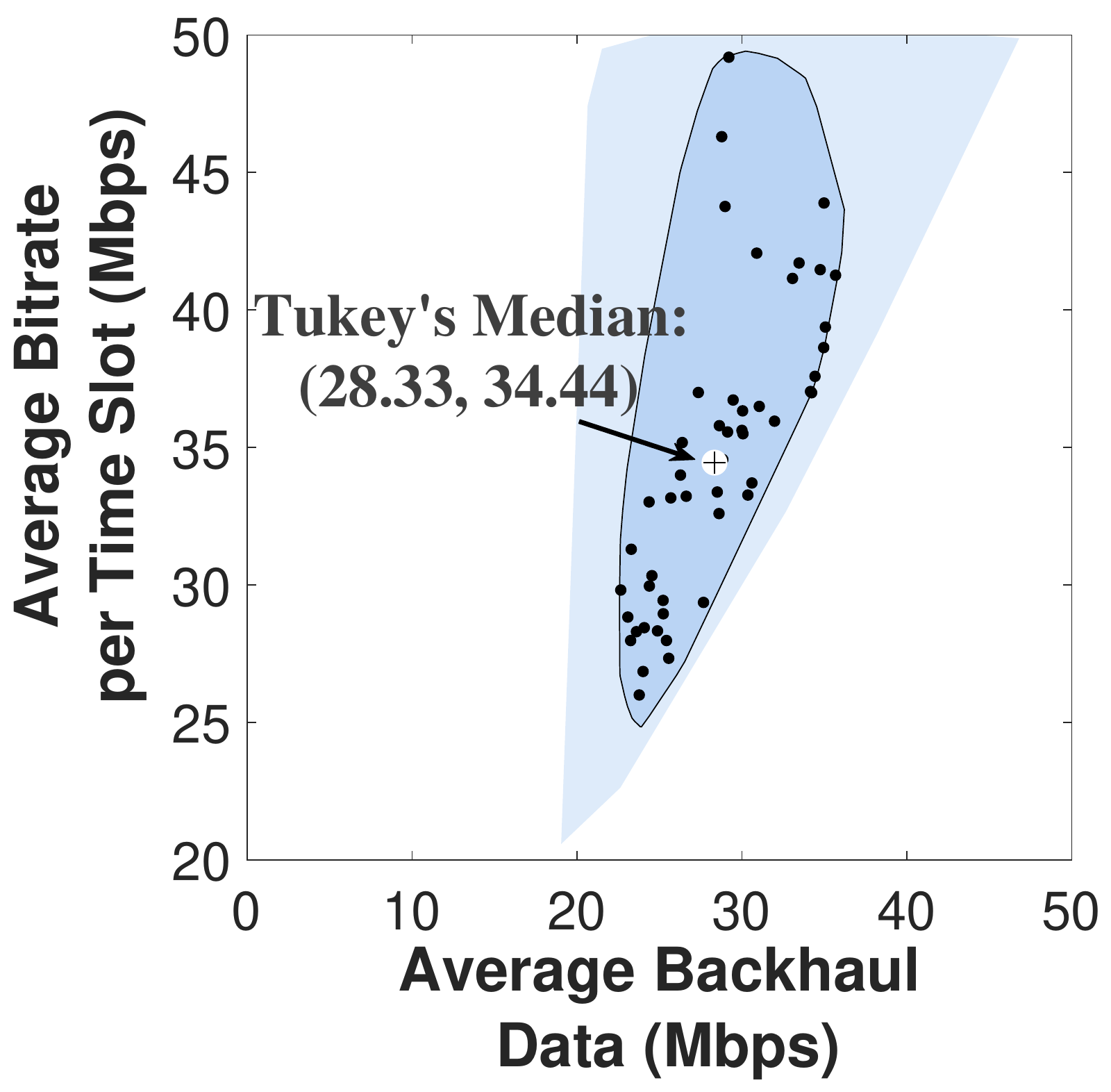} 
    \caption{Joint QoE-traffic} 
    \label{fig:bagplotjointqoetrafficfixed}
    \vspace{0.32ex}
  \end{subfigure}
  \begin{subfigure}[b]{0.33\linewidth}
    \centering
   \includegraphics[width=1\linewidth]{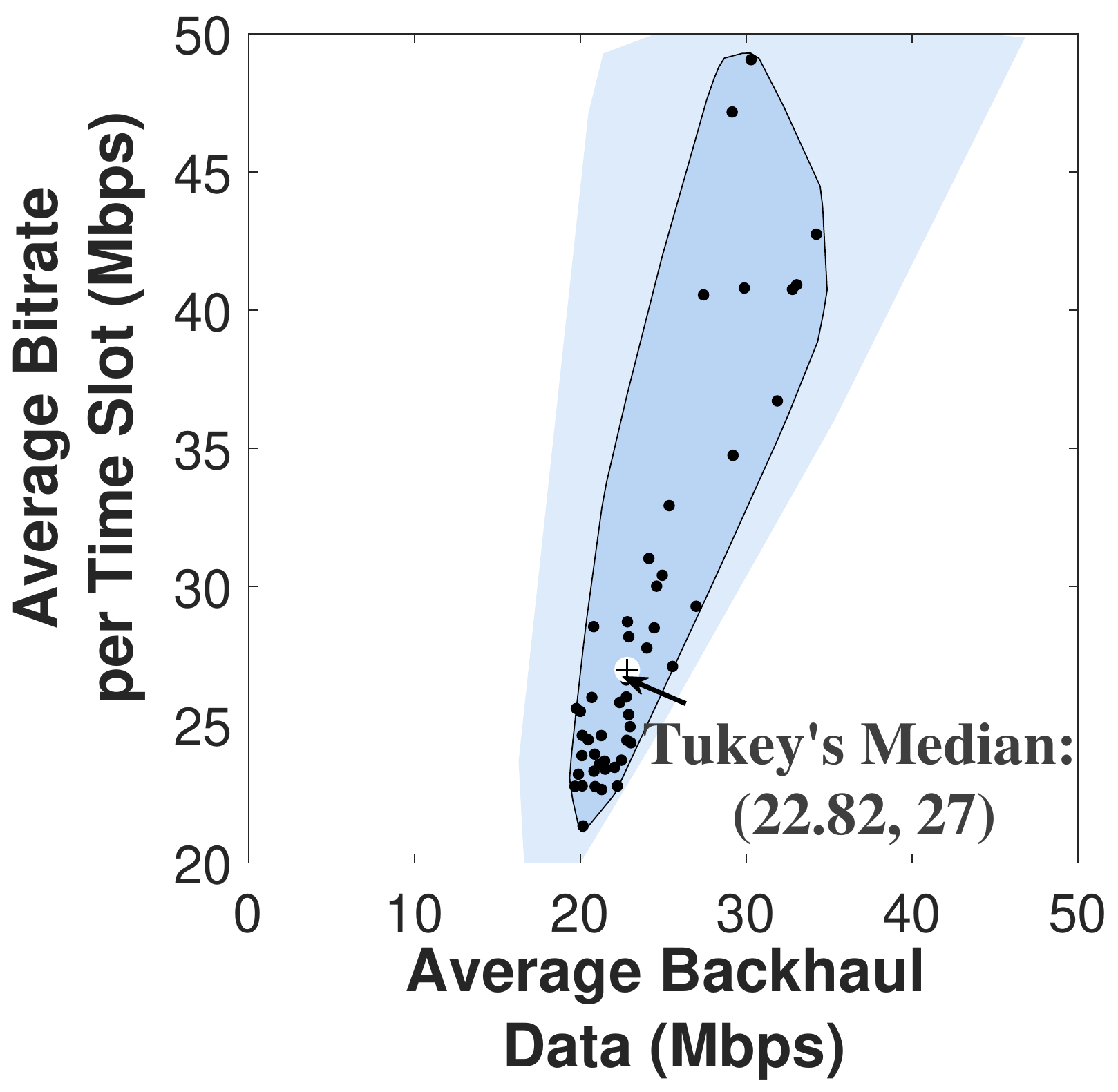} 
    \caption{Data Traffic} 
    \label{fig:bagplotdatatrafficfixed}
    \vspace{0.12ex}
  \end{subfigure} 
  \begin{subfigure}[b]{0.33\linewidth}
    \centering
    \includegraphics[width=1\linewidth]{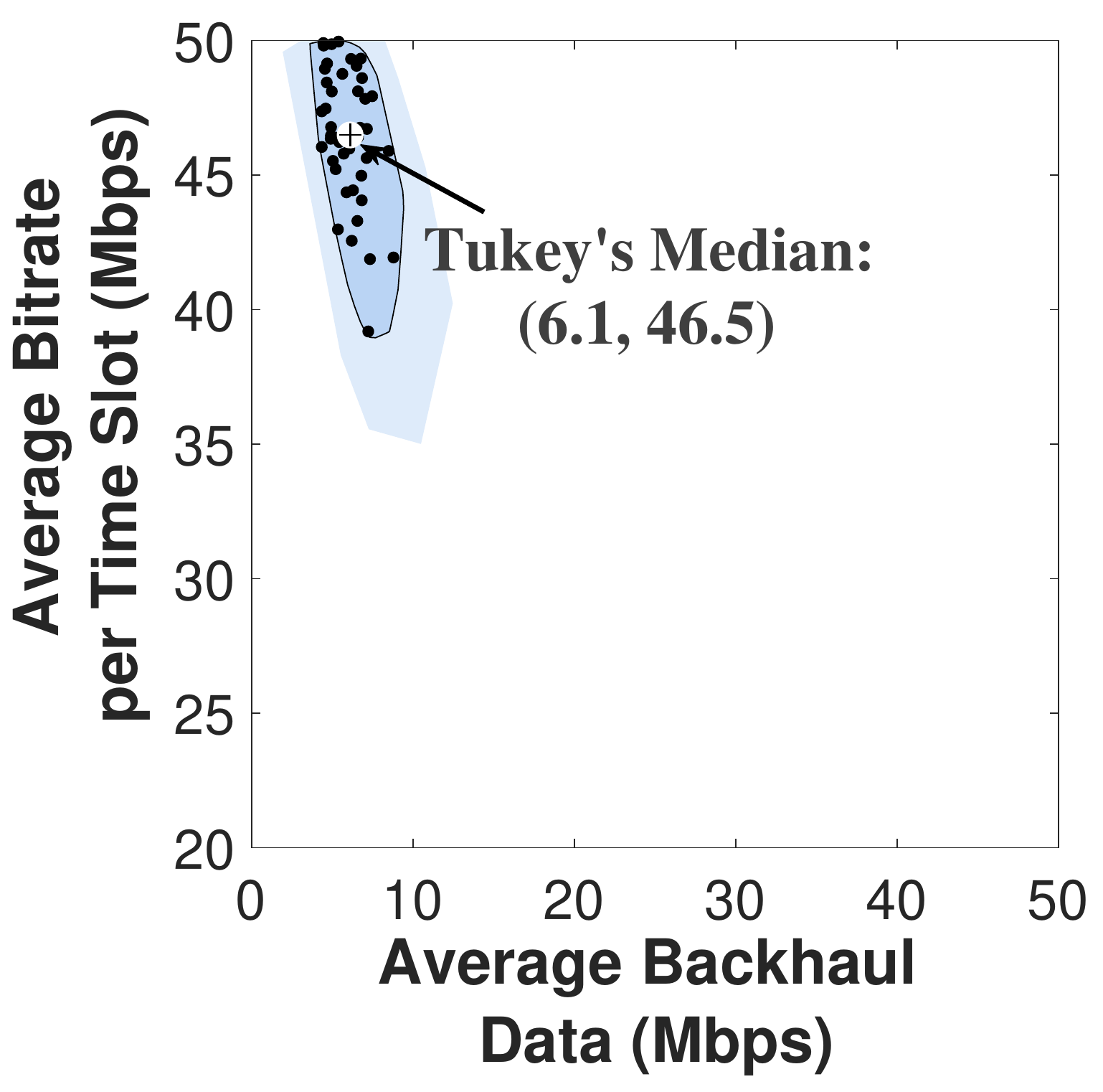} 
    \caption{QoE} 
    \label{fig:bagplotqoecache}
    \vspace{0.32ex}
  \end{subfigure}
  \begin{subfigure}[b]{0.33\linewidth}
    \centering
    \includegraphics[width=1\linewidth]{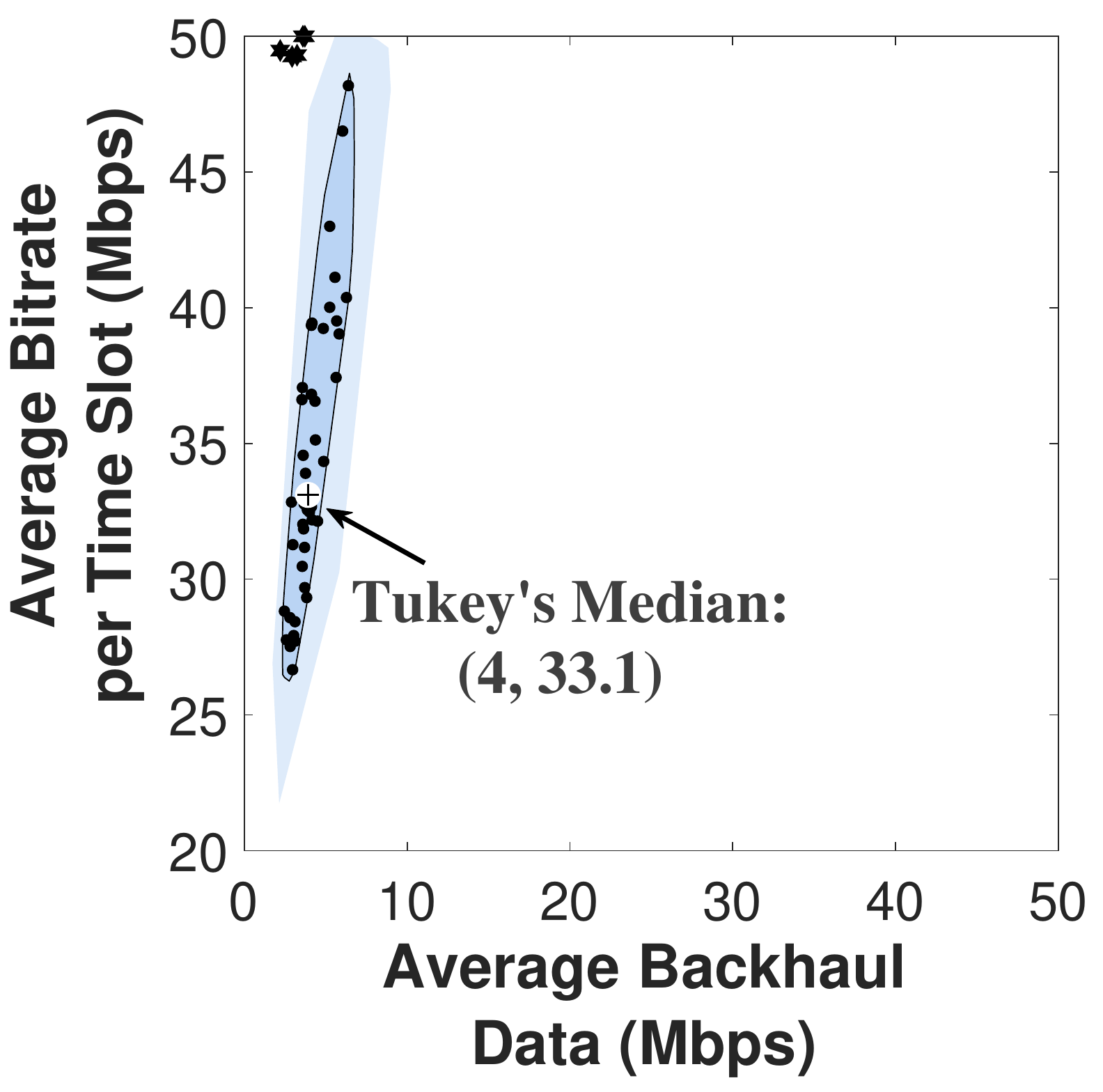}  
    \caption{Joint QoE-traffic} 
    \label{fig:bagplotjointqoetrafficcache}
    \vspace{0.32ex}
  \end{subfigure}
  \begin{subfigure}[b]{0.33\linewidth}
    \centering
    \includegraphics[width=1\linewidth]{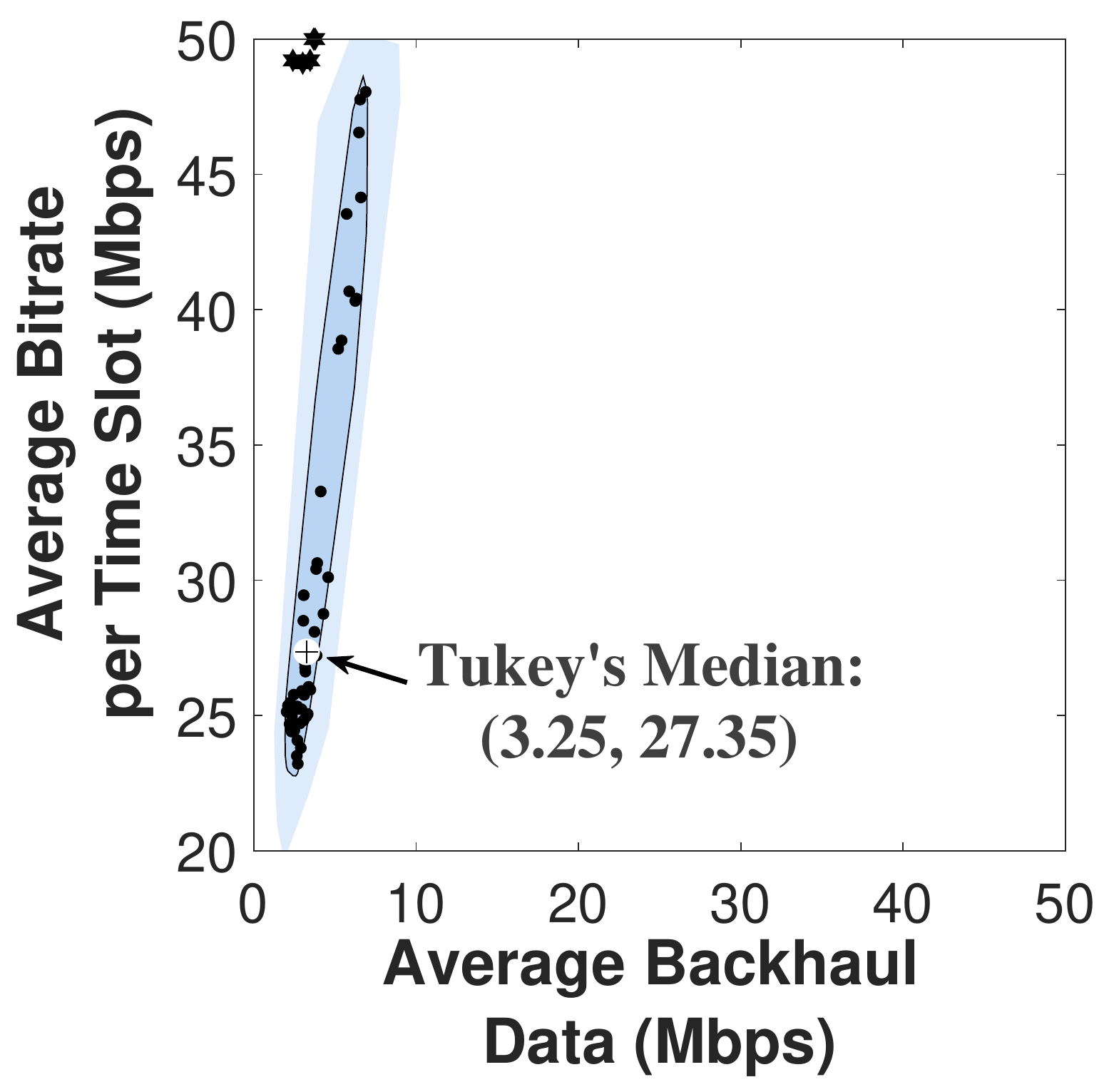}  
    \caption{Data Traffic} 
    \label{fig:bagplotdatatrafficcache}
    \vspace{0.32ex}
  \end{subfigure}
  \caption{Average bitrate/data traffic bag plots and half space Tukey's median for the case of fixed contents in the cache (a-c) and the case of cache updating (d-f).}  
  \label{fig:bagplots}
\end{figure}

In Figure \ref{fig:bagplots}, we visualize the bitrate-traffic trade-off of the three strategies under the two different scenarios using the bag plots with Tukey's half space median \cite{BagPlot}. For drawing the two-dimensional bag plots, we use the LIBRA Matlab libraries \cite{LIBRA}. Fig. \ref{fig:bagplotqoefixed}-\ref{fig:bagplotdatatrafficfixed} shows the bagplots under the fixed cached contents scenario. The data traffic minimization strategy localizes the access of most of the clients to the edge servers and therefore reduces the backhaul traffic. However, in this case, most of the clients suffer from the low average bitrate compared to other optimization strategies. In contrast, QoE maximization strategy tends to achieve the maximum bitrates for the clients while generating significant backhaul traffic. 

Figures \ref{fig:bagplotqoecache}-\ref{fig:bagplotdatatrafficcache} contain the bag plots corresponding to the scenario with dynamic cache contents. The plots visually tell us the same story that regardless of the optimization strategy used a substantial reduction in backhaul traffic is achieved with dynamic cache contents compared to fixed caches. Another interesting observation from the plots in Fig. \ref{fig:bagplotqoefixed}-\ref{fig:bagplotdatatrafficcache} is that the deviation of the created traffic by the clients on the origin server from the average one is less (narrow bag) under the cache updating compared to the fixed cached contents scenario. This observation suggests that with cache updating, the clients download from the origin server causes less and more smooth data traffic on the backhaul network. For each individual client, this smooth behavior of data traffic can be also observed from the plots in Fig. \ref{fig:averagedata} and \ref{fig:averagedatacache}.     



\subsection{Comparing Different Cache Replacement Heuristics}
\label{subsec:cacheupdatingheuristics}

In this section, we compare the performance of the proposed proactive cache replacement heuristic RBCRH with two popular cache updating approaches which are least recently used (LRU) and least frequently used (LFU). The joint QoE-traffic optimization strategy with the weighting $\beta=0.5$ is employed for the following simulations. The heuristics are compared in term of the percentage of cache miss per client and chunk i.e. the ratio between the missed chunks in the cache to the accessed chunks during the whole video streaming of all clients. We also compare three approaches with the offline optimal cache replacement strategy. Knowing in advance the one slot ahead status of the clients (departure time slot, throughput), the optimal approach at each time slot caches the chunks that minimize cache miss in the next time slot.  

We should note that the proposed cache replacement method described in \cite{Pedersen2016} is a variation of LRU which in contrast to our heuristic does not take into account the retention of the clients during different parts of the video. We compare our heuristic with original LRU and the results remain also valid for its proposed variation in \cite{Pedersen2016}.

\begin{figure}[t] 
  \begin{subfigure}[b]{0.49\linewidth}
    \centering
     \includegraphics[width=1\linewidth]{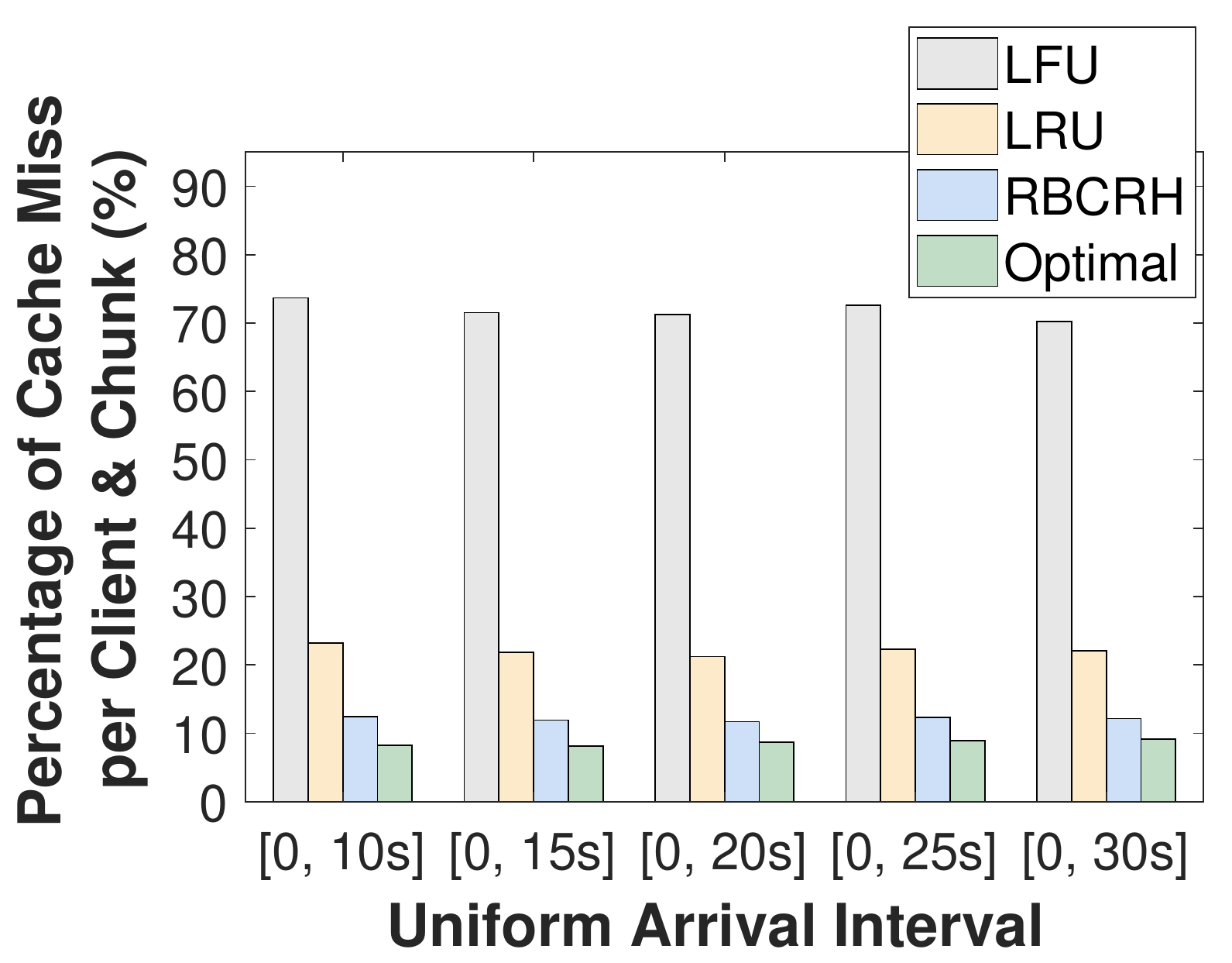}  
    \caption{Effect of Arrival Interval} 
    \label{fig:cachereplacementarrival}
    \vspace{0.2ex}
  \end{subfigure}
  \begin{subfigure}[b]{0.48\linewidth}
     \centering
     \includegraphics[width=1\linewidth]{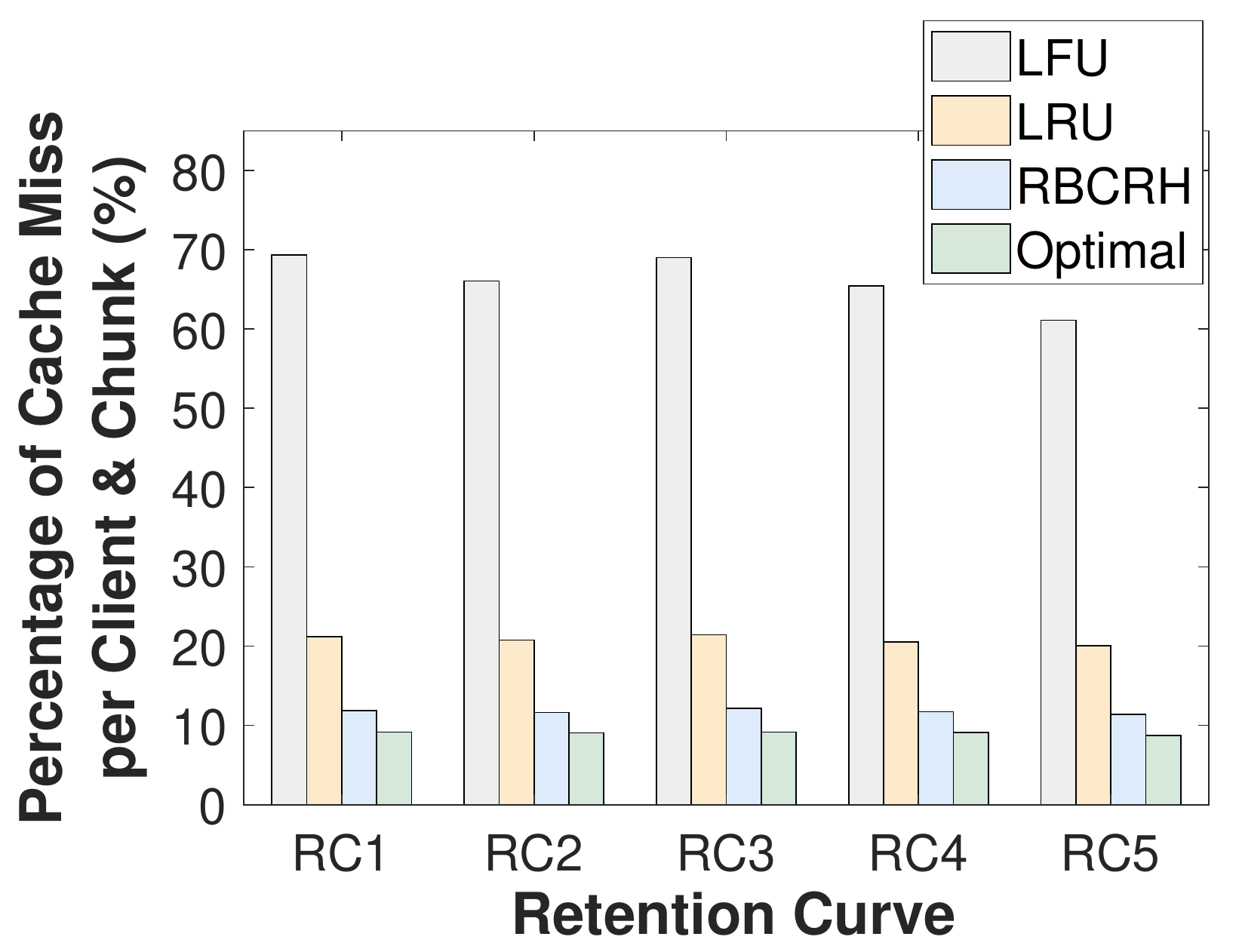} 
     \caption{Effect of Retention Curve} 
     \label{fig:cachereplacementretention}
     \vspace{0.32ex}
  \end{subfigure} 
  \caption{Comparison between cache replacement heuristics for different a) arrival intervals b) retention curves.}  
\end{figure}

\subsubsection{Clients Arrival Dynamics}

Fig. \ref{fig:cachereplacementarrival} shows the comparison result considering different uniform arrival intervals of the mobile clients. We observe that using RBCRH which takes into account two statistical indexes accessibility history and the retention estimation of the clients, yields noticeable reduction in the percentage of cache miss compared to both LRU and LFU. For different arrival intervals, RBCRH heuristic reduces the cache miss percentage by in average about 83\% and 45\% compared to LFU and LRU, respectively.  
Concerning the optimal approach, interestingly, the percentage of cache miss using our heuristic is in average less than 1.4 times of the optimal cache miss for different arrival intervals, hence, providing the average approximation factor of 1.4 for the cache miss minimization. It is noteworthy to mention that the gap in the percentage of cache miss between RBCRH and the optimal strategy is due to the fact that our heuristic works online by relying only on the statistical information of the clients without knowledge of their future status.

\subsubsection{Robustness} 

We have also evaluated the robustness of the proposed cache replacement heuristic by considering different retention behaviors of the mobile clients. Clients arrive within the interval $[0,30s]$ and they stay active for maximum 270 seconds. We use the linear quadratic polynomial $p(t)=at^2+bt+c$ to generate the retention curve (RC) which describes the probability that the client leaves its streaming session at time slot $t$. Different retention curves are generated by changing the curvature of polynomial $p(t)$ and determining the corresponding coefficients $a$, $b$ and $c$. The comparative results for five different retention curves RC1$-$RC5 are plotted in Fig. \ref{fig:cachereplacementretention}. Note that here the sharpness of the polynomial increases as the curve index increases, for instance, the probability that a client remains active at given time slot $t$ under retention curve RC5 is less than that for curve RC4.    

The results show that the proposed heuristic RBCRH outperforms both LRU and LFU in term of the percentage of cache miss per chunk for different retention behaviors as well. For this case, RBCRH reduces the cache miss percentage by average about 82\% and 43\% compared to LFU and LRU, respectively.
For different retention curves, the percentage of cache miss using our heuristic is again not more than 1.4 times of the optimal cache miss in average. Similar to the case of arrival interval, the gap in the percentage of cache miss between our heuristic and the optimal one is due to the lack of exact knowledge in our heuristic about the future status of the clients.

\begin{figure}[t] 
  \begin{subfigure}[b]{0.43\linewidth}
    \centering
     \includegraphics[width=1\linewidth]{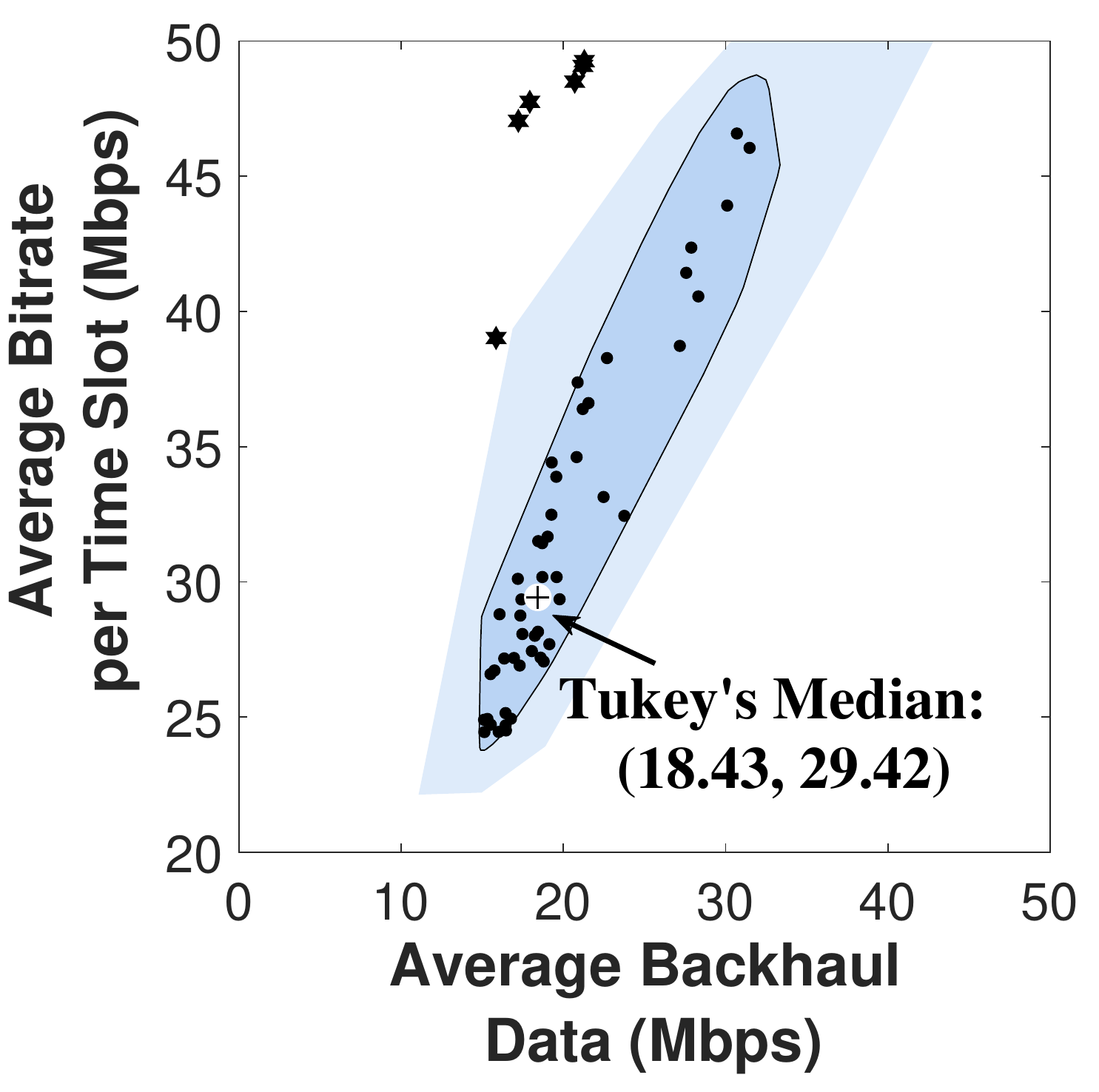} 
    \caption{LFU} 
    \label{fig:bagplotlfu}
    \vspace{0.2ex}
  \end{subfigure}
  \begin{subfigure}[b]{0.43\linewidth}
    \centering
   \includegraphics[width=1\linewidth]{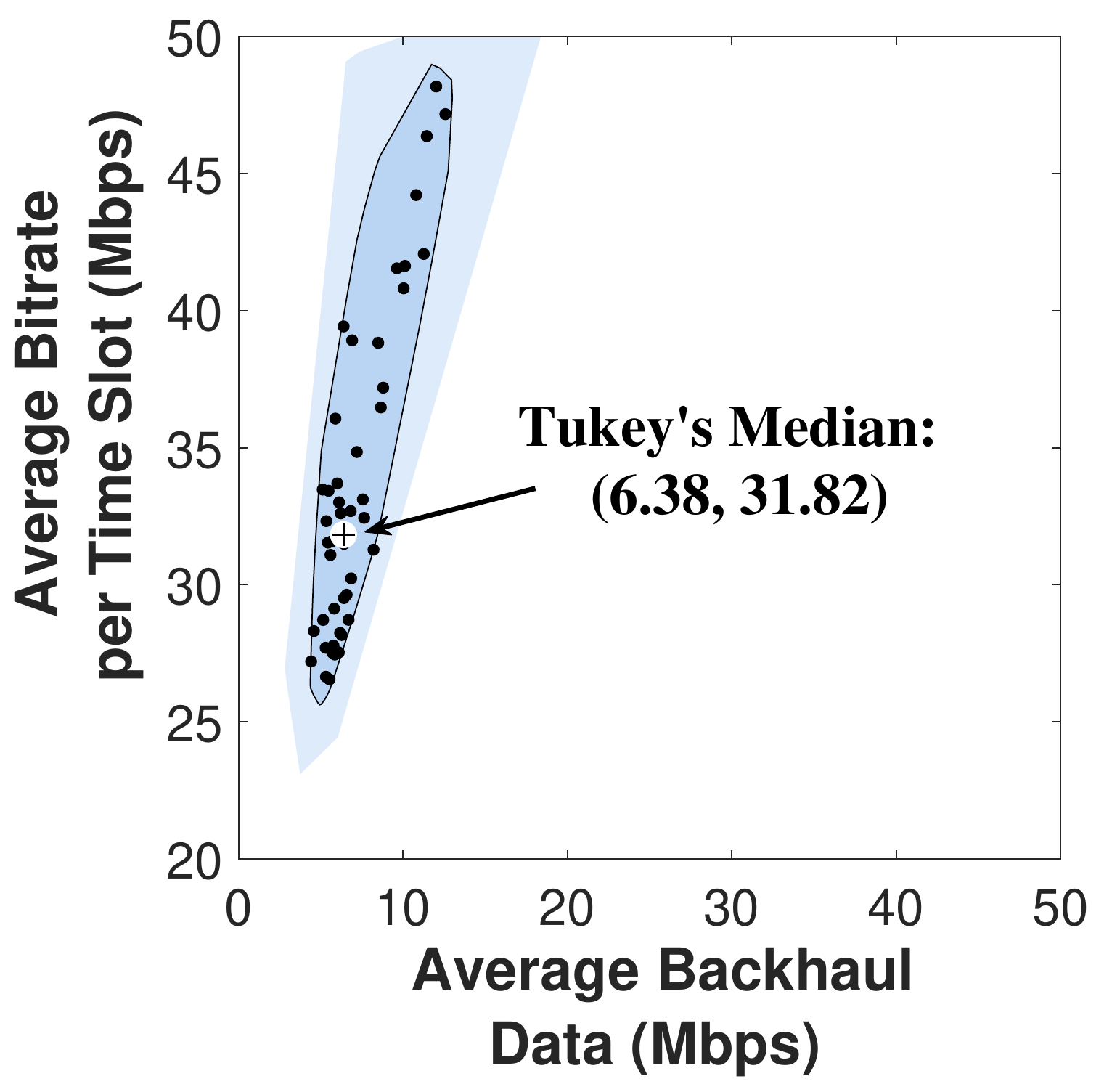} 
    \caption{LRU} 
    \label{fig:bagplotlru}
    \vspace{0.32ex}
  \end{subfigure}    
  \begin{subfigure}[b]{0.43\linewidth}
    \centering
   \includegraphics[width=1\linewidth]{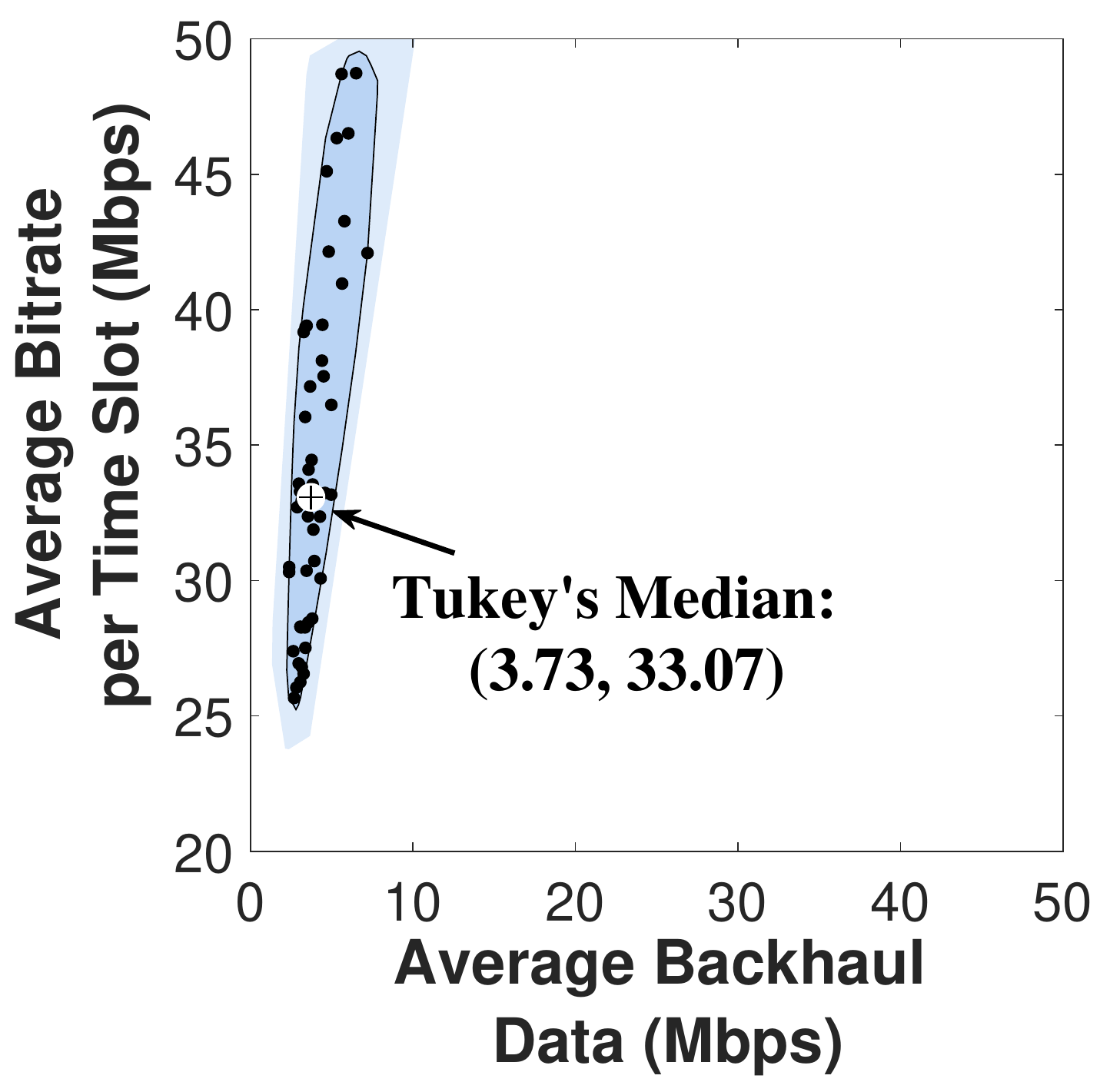} 
    \caption{RBCRH} 
    \label{fig:bagplotrbcrh}  
    \vspace{0.12ex}
  \end{subfigure}
  \begin{subfigure}[b]{0.43\linewidth}
    \centering
   \includegraphics[width=1\linewidth]{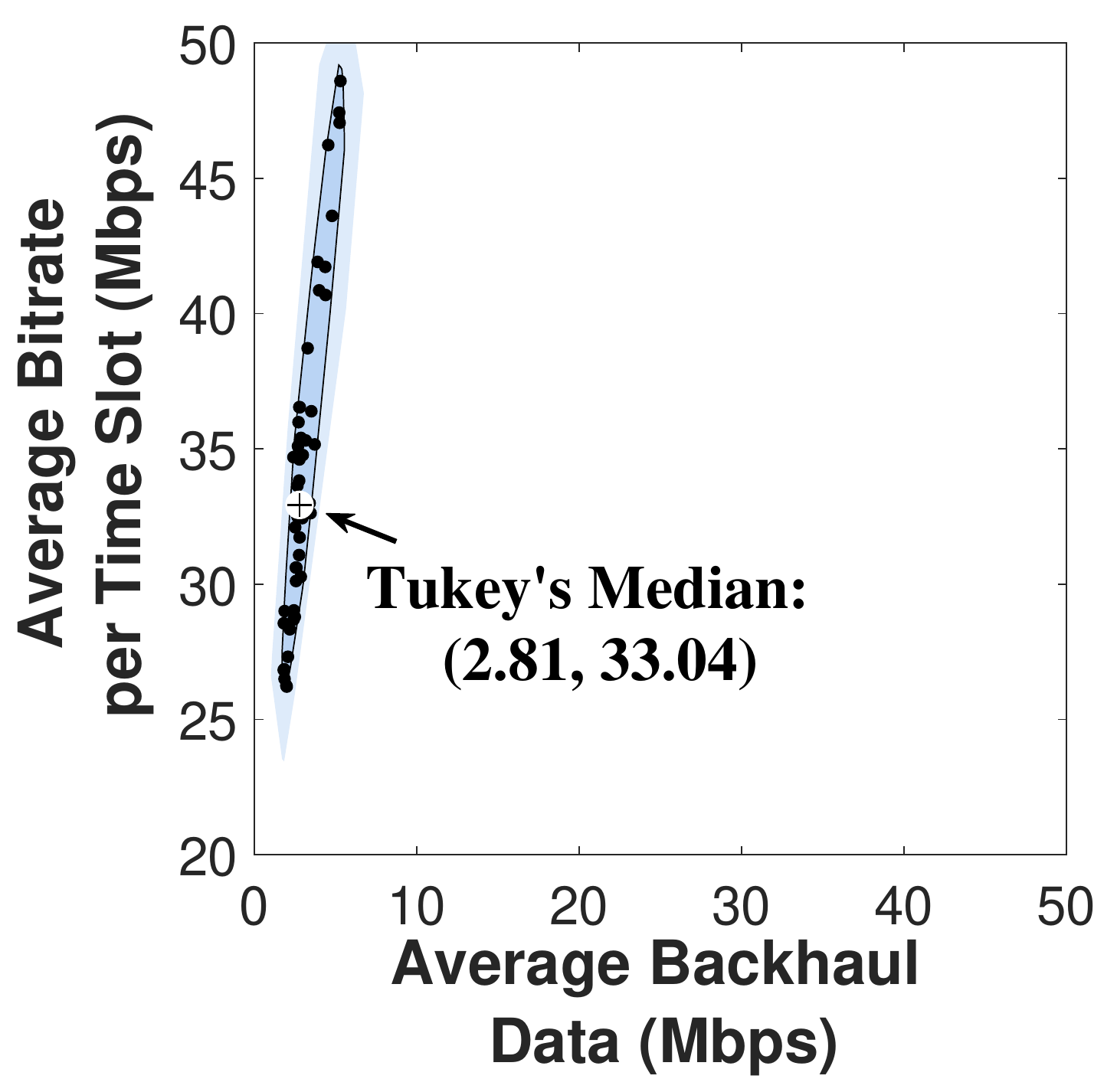} 
    \caption{Optimal} 
    \label{fig:bagplotoptimal}
    \vspace{0.12ex}
  \end{subfigure} 
  \caption{Average bitrate/data traffic bag plots and Tukey's half space median for different cache replacement heuristics.}  
\end{figure}

\subsubsection{QoE and Backhaul Traffic Comparison}

With arrival interval $[0,30s]$ and linear retention curve, we have also shown in Fig. \ref{fig:bagplotlru}-\ref{fig:bagplotoptimal} the bag plot of average bitrate and backhaul data traffic together with the half space Tukey's median for the cache replacement heuristics. As we can see from the result, using RBCRH, the average data traffic reduces compared to LRU while achieving higher average birate. This obviously shows the superiority of the proposed heuristic compared to both LRU and LFU. It is also noticed from the results that under adaptive multiple video streaming scenario, LFU heuristic shows lower performance compared to LRU.   

Fig. \ref{fig:bagplotoptimal} shows that, as expected, with the exact knowledge of the future status of mobile clients using the optimal strategy, the lower data traffic is created on the origin server compared to our heuristic. This is because with the optimal strategy, the clients fetch their desired chunks in the most cases from the edge cache.     
Another observation from the plots is that the difference in average data traffic between the heuristics is more than the different in average bitrate. Similar to our previous observation, this in turn confirms that the impact of cache replacement using different heuristics is more on the backhaul data traffic than the QoE of the clients.

\section{Conclusion}
\label{sec:conclusion}

In this work, we have investigated the impact of video caching and replacement on the optimal QoE of the mobile clients in edge caching mobile adaptive video streaming scenarios. 
We proposed a weighted optimization problem for jointly maximizing the QoE of individual DASH clients, the fair bitrate allocation as well as minimizing the overall data traffic on the origin server. We then designed an online greedy-based algorithm for solving the optimization problem which is run by the coordinator. The low complexity of the algorithm and minimum need for parameter tuning makes it feasible for easy deployment by the mobile network operators (MNOs). An effective cache updating heuristic is also proposed which takes into account not only the content popularity according to the past history but also the retention pattern of the clients during different parts of the video. The proposed cache replacement strategy reduces significantly the percentage of cache miss compared to the alternative strategies and shows very close performance to the offline optimal solution.   

The results of our conducted simulations reveal that with fixed contents cached at the edges, the impact of video caching on the optimal QoE of the clients is proportional to the desired operational point of MNO. Furthermore, in circumstances with constrained cache size, the effect of cache replacement on the QoE is less noticeable compared to its effect on the volume of data downloaded from the origin server.  

As the future works, we are planning to exploit the potential of collaborative edge caching in joint QoE-traffic optimization in which the requested chunks can be delivered from not only the local but also the neighborhood edge servers within a cluster.


%

\section*{Acknowledgment}

This work has been financially supported by the Academy of Finland (grant number 297892), Tekes - the Finnish Funding Agency for Innovation, and the Nokia Center for Advanced Research.

\ifCLASSOPTIONcaptionsoff
  \newpage
\fi

\end{document}